\documentclass[fleqn,usenatbib]{mnras}

\usepackage{newtxtext,newtxmath}

\usepackage[T1]{fontenc}

\DeclareRobustCommand{\VAN}[3]{#2}
\let\VANthebibliography\thebibliography
\def\thebibliography{\DeclareRobustCommand{\VAN}[3]{##3}\VANthebibliography}

\usepackage{graphicx}	
\usepackage{amsmath}	
\usepackage{subcaption}
\usepackage{ulem, cancel}

\title[Shannon entropy and the stability of Kepler-60]{Application of the Shannon entropy in the planar (non-restricted) four-body problem: the long-term stability of the Kepler-60 exoplanetary system}

\author[K\H ov\'ari, \'Erdi, \& S\'andor]{
E. K\H ov\'ari$^{1,2}$\thanks{E-mail: E.Kovari@astro.elte.hu},
B. \'Erdi$^{1,2}$\thanks{E-mail: B.Erdi@astro.elte.hu},
Zs. S\'andor$^{1,2,3}$\thanks{E-mail: Zs.Sandor@astro.elte.hu}\\
$^1$Department of Astronomy, Institute for Geography and Earth Sciences, E\"otv\"os Lor\'and University, P\'azm\'any P\'eter s\'et\'any 1/A, H-1117 Budapest, Hungary\\
$^2$Centre for Astrophysics and Space Science, E\"otv\"os Lor\'and University, P\'azm\'any P\'eter s\'et\'any 1/A, H-1117 Budapest, Hungary\\
$^3$Konkoly Observatory, Research Centre for Astronomy and Earth Sciences, Konkoly Thege Mikl\'os \'ut 15-17., H-1121 Budapest, Hungary
}
\date{Accepted 2021 October 8. Received 2021 October 8; in original form 2020 December 15.}

\pubyear{2021}

\begin{document}
\label{firstpage}
\pagerange{\pageref{firstpage}--\pageref{lastpage}}
\maketitle

\begin{abstract}
In this paper, we present an application of the Shannon entropy in the case of the planar (non-restricted) four-body problem. Specifically, the Kepler-60 extrasolar system is being investigated with a primary interest in the resonant configuration of the planets that exhibit a chain of mean-motion commensurabilities with the ratios 5:4:3. In the dynamical maps provided, the Shannon entropy is utilized to explore the general structure of the phase space, while, based on the time evolution of the entropy, we determine also the extent and rate of the chaotic diffusion as well as the characteristic times of stability for the planets. Two cases are considered: (i) the pure Laplace resonance when the critical angles of the $ 2 $-body resonances circulate and that of the $ 3 $-body resonance librates; and (ii) the chain of two $ 2 $-body resonances when all the critical angles librate. Our results suggest that case (ii) is the more favourable configuration but we state too that, in either case, the relevant resonance plays an important role to stabilize the system. The derived stability times are no shorter than $ 10^{8} $~yrs in the central parts of the resonances.
\end{abstract}

\begin{keywords}
celestial mechanics -- chaos -- diffusion -- planets and satellites: dynamical evolution and stability
\end{keywords}



\section{Introduction}
\label{sec:introduction}

Investigating the overall dynamics of resonant planetary systems is a challenging problem in celestial mechanics. In the vicinity of mean-motion resonances (MMRs), the phase space reveals a wide range of interesting phenomena: stable islands of quasi-periodic motion and thin chaotic layers attached to their boundaries, the overlap of neighbouring resonances in the case of larger perturbations acting and hence the appearance of more extended chaotic domains, the birth of secondary (tertiary, etc.) resonances, and so forth. The presence of secular resonances further enhances the complexity of dynamics. Such variety of phenomena makes the evolution of the individual planets - as well as that of the whole system - far from being straightforward, and it also showcases the demanding yet important nature of dynamical studies.

The stability of a dynamical system might be defined in numerous different ways. The concepts of the Lyapunov, Lagrangian, Hill, Arnold, AMD (etc.) stability are all widely used in planetary dynamics. Most commonly, the definition of the Lyapunov stability is adopted, according to which an orbit of a dynamical system is stable if nearby, i.e. perturbed, orbits remain in its neighbourhood. Based on this definition, the chaoticity can be quantified by calculating the Lyapunov Characteristic Exponents (LCEs) of a trajectory. Intriguing is the case, however, of the so-called 'stable chaos' \citep[see e.g.][]{milani1992} where notwithstanding the local exponential divergence of close-by orbits (i.e. positive LCEs), the chaos in phase space does not lead to physical instabilities and thus the trajectory in question exhibits regular motion for timescales much longer than its Lyapunov time. The key quantity in such cases is the chaotic diffusion for its extent will determine the evolution and long-term dynamics of a system.

Although the estimation of the chaotic diffusion was studied in several papers already \citep[see e.g.][]{froeschle2005,lega2007,cincotta2018}, the calculations are very time-consuming and require rather long integrations. This is why the recent applications of the Shannon entropy \citep{shannon1949} might be considered as a breakthrough in planetary dynamics. \cite{cincotta2012}, \cite{giordano2018}, and \cite{cincotta2020} showed that the Shannon entropy provides a useful numerical tool to measure the extent of unstable regions in action space as well as to estimate the rate of the chaotic diffusion. These early applications of the entropy were limited to discrete dynamical maps of low dimensions, whereas the first application to a more complex, planetary-type problem was presented in \cite{beauge2019} where the dynamics of the planar restricted three-body problem (3BP) was investigated in the vicinity of first-order mean-motion resonances. Later, \cite{cincotta2021febr} and \cite{cincotta2021march} studied the planar (non-restricted) 3BP, more specifically, they have chosen the HD 20003 and HD 181433 exoplanetary systems, respectively, to test the sensitivity of the Shannon entropy formalism on real systems. In the present paper, we extend the use of the Shannon entropy to the planar (non-restricted) four-body problem (4BP) via performing a dynamical analysis of the Kepler-60 three-planet extrasolar system.

The paper is organized as follows. In Section~\ref{sec:the_kepler-60_exoplanetary_system}, we introduce the Kepler-60 planetary system with two types of resonant configurations that might govern the dynamics of the planets. A brief summary of the latest results on the Shannon entropy, as well as its numerical computation and that of the diffusion coefficient and stability times, is given in Section~\ref{sec:the_shannon_entropy_formalism}. Section~\ref{sec:results_and_discussion} is devoted to the discussion of our results on the Kepler-60 system. In this section we compute the Shannon entropy, the diffusion coefficients, and the characteristic times of stability in the proximity of each planet in the phase space. The two types of resonant configurations are studied separately. We also compare our results with direct, long-term numerical integrations. A summary of the results is given in Section~\ref{sec:summary}.

\section{The Kepler-60 exoplanetary system}
\label{sec:the_kepler-60_exoplanetary_system}

The Kepler-60 system consists of three super-Earth planets (nominated as Kepler-60$b $, $c $, and $ d $) orbiting a central star of type G, of mass $ 1.105 m_\odot $, and of radius $ 1.448 R_\odot $ \citep{rowe2015}. With the orbital periods of the $ \sim 4 m_{\oplus} $ planets being close to $ \sim 7 $, $ \sim 9 $, and $ \sim 12 $ days \citep{steffen2013}, the configuration appears to be an extremely compact one. This property itself promises interesting dynamics. Moreover, the mean-motion ratios of the subsequent planets are very close to the commensurabilities 5:4 and 4:3, suggesting a pair of 2-body MMRs operating, but also, a specific linear combination of all three of the mean motions indicates the presence of a $ 3 $-body resonance, too, with the ratios 5:4:3. Such rich combination of 2- and 3-body resonances with potential overlaps, further enhances the interest toward the system.

The critical angles for the $ 2 $-body resonances of the consecutive pairs can be formulated as
\begin{align}
    \Phi_1 &= (p+1) \lambda_c - p\lambda_b - \widetilde{\omega}_b, \label{eq:fi1}\\
    \Phi_2 &= (p+1) \lambda_c - p\lambda_b - \widetilde{\omega}_c, \label{eq:fi2}\\
    \Phi_3 &= (q+1) \lambda_d - q\lambda_c - \widetilde{\omega}_c, \label{eq:fi3}\\
    \Phi_4 &= (q+1) \lambda_d - q\lambda_c - \widetilde{\omega}_d, \label{eq:fi4}
\end{align}
where $ p, q > 0 $ are prime integers that define the first-order resonances $ \frac{p+1}{p}=\frac{5}{4} $ and $ \frac{q+1}{q} = \frac{4}{3} $ of the inner and outer pairs of planets, respectively; $ \lambda_i $ ($ i = b,c,d $) are the mean longitudes; and $ \widetilde{\omega}_i $ ($ i = b,c,d $) are the pericentre longitudes of the planets (with $ i = b $ corresponding to the inner- and $ i = d $ to the outermost planet). As for the $ 3 $-body MMR, it can be regarded as a generalized Laplace resonance where the Laplacian period ratios 2:1 are replaced by the first-order resonance ratios 5:4 and 4:3 \citep[see][]{papaloizou2015}. The critical angle for this resonance is
\begin{equation}
    \Phi_{\mathrm{L}} = -p\lambda_b + (p+q+1)\lambda_c - (q+1)\lambda_d,
    \label{eq:fiL}
\end{equation}
where the connotation of the letters is the same as above.

From simply a mathematical point of view - and with the assumption that the $ 3 $-body (or Laplace) resonance is fulfilled, i.e., $ \Phi_{\mathrm{L}} $ librates, one can distinguish between 5 cases regarding the librating-circulating nature of the critical angles of the $ 2 $-body resonances: (i) none of the critical angles \eqref{eq:fi1}-\eqref{eq:fi4} librate; (ii) all four of them librate; (iii) one of them librates; (iv) two of them librate; (v) three of them librate. (We note that in case (ii), the libration of $ \Phi_{\mathrm{L}} $ is a consequence; therefore, it holds even without having to assume it. We also note that - if not in the framework of the Laplace resonance - one could add $ 3 $ more cases: the cases (iii), (iv), and (v) without $ \Phi_{\mathrm{L}} $ librating. (Case (i) with a circulating $ \Phi_{\mathrm{L}} $, i.e., a non-resonant system is irrelevant in our case since it contradicts the observational data.))

Case (v) with $ \Phi_{\mathrm{L}} $ circulating and case (ii) were studied by \cite{papaloizou2015} who came to the conclusion that the most likely scenario is that - after a convergent planet migration induced by interactions with the protoplanetary disc - the system was formed in a state of resonant chain (case (ii)). However, he also found that this case implies that the resonant period ratios increasingly depart from the strict commensurabilities when evolving in time.
 
\cite{gozdziewski2016} were looking for best-fitting solutions of the original TTV data \citep{steffen2013} of the Kepler-60 system and proposed two possible solutions of resonant configurations, corresponding to the cases (i) and (ii). (Following their terminology, let us refer, henceforth, to case (i) as the 'pure Laplace resonance'.) Similarly to \cite{papaloizou2015}, the authors also argue that a convergent migration leads to the formation of a chain of 2-body MMRs (case (ii)), thus they claim that the pure Laplace resonance seems unexpected. 

Nonetheless, a dynamical analysis carried out by \cite{gozdziewski2016} - and later by \cite{panichi2017} as well - made it clear that the system is dynamically particularly active. Due to the overlap of $ 2 $- and $ 3 $-body resonances, the phase space is highly chaotic, the zones of stable motion are confined to isolated islands of the MMRs, and the structure of the Arnold web also appears in the dynamical maps.

Thus the Kepler-60 system serves as an excellent candidate to test the Shannon entropy technique on a resonant four-body system. By directly computing the chaotic diffusion and the stability times of the planets, we hope to shed some more light on the yet ambiguous dynamics of this intriguing system.

\section{The Shannon entropy formalism}
\label{sec:the_shannon_entropy_formalism}

In this section, we briefly summarize the formalism of the Shannon entropy approach, following the works of \cite{giordano2018}, \cite{beauge2019}, and \cite{cincotta2021febr,cincotta2021march}, among others.

Let us consider, for simplicity but without the loss of generality, a 2-dimensional discrete dynamical system with action variables $ (I_1, I_2) \in \mathcal{B} \subset \mathbb{R}^2 $ and angle variables $ (\vartheta_1, \vartheta_2) \in \mathbb{S}^1\times \mathbb{S}^1 $ ($ \mathcal{B} $ is some open domain of the plane and $ \mathbb{S}^1 $ is the unit circle). Let then $ M $ denote the 4-dimensional map $ (I_1, I_2, \vartheta_1, \vartheta_2)(t_i) \to (I_1, I_2, \vartheta_1, \vartheta_2)(t_{i+1}) $ ($ i=1, \dots, \infty $). By fixing some values $ \vartheta_1^0 $ and $ \vartheta_2^0 $ of the phases, one can define a 2D section of the full 4D map $ M $ as
\begin{equation}
    \mathcal{S} = \{\left(I_1, I_2\right)(t_i), \ i=1, \dots \infty: |\vartheta_1-\vartheta_1^0|+|\vartheta_2-\vartheta_2^0|<\delta \ll 1\}.
    \label{eq:mathcalS}
\end{equation}
Introduce then a partition
\begin{equation}
    \alpha = \{\beta_i: i = 1, \dots, r\} \subset \mathcal{S}
    \label{eq:alpha}
\end{equation}
such that its $ r $ bi-dimensional elements $ \beta_i $ cover the whole $ \mathcal{S} $ and are disjoint and measurable. Let
\begin{equation}
    \gamma \equiv \gamma_I = \{\left(I_1, I_2\right)(t_i), \ i=1, \dots, \infty\}
    \label{eq:gamma_I}
\end{equation}
denote the projection of the full trajectory
\begin{equation}
    \gamma_{(I, \vartheta)} = \{\left(I_1, I_2, \vartheta_1, \vartheta_2)\right)(t_i), \ i=1, \dots, \infty\}
    \label{eq:gamma_Itheta}
\end{equation}
to the action plane, and let $ N \equiv N(t) $ be the number of intersections of $ \gamma $ with the section $ \mathcal{S} $ (until some time $ t $). If we denote with $ n_k \equiv n_k(t) $ the number of intersecting points that fall in the $ k $-th cell of the partition (again, until some time $ t $), then the Shannon entropy of $ \gamma $ for the partition $ \alpha $ is given by
\begin{equation}
    S(t; \gamma, \alpha) = \ln(N) - \frac{1}{N} \sum_{k=1}^{r} n_k \ln(n_k).
    \label{eq:S}
\end{equation}

It can easily be shown that for any given partition and trajectory $ S $ is always bounded: $ 0\leq S(t; \gamma, \alpha) \leq \ln(r) $. The minimal value 0 characterizes the case of perfect stability, i.e., when all the intersections of $ \gamma $ with $ \mathcal{S} $ are in one single cell of the partition ($ n_k = \delta_{ik} $ is the Dirac delta), whereas the maximum $ \ln(r) $ is reached when the motion is completely ergodic and the elements of $ \alpha $ are filled uniformly ($ n_k = N/r $). To apply the Shannon entropy as an indicator of chaos, it is convenient to take advantage of the above property of boundedness and use the normalized entropy $ S / \ln(r) \in [0, 1] $ to measure the stability of a given orbit in the following way: for regular orbits, this quantity rapidly reaches a constant value, which is significantly smaller than the maximal $ 1 $ (this means that both the number of the occupied cells and the distribution of $ n_k $ within them is stabilized quickly, indicating a well-confined region in action space being covered by the trajectory), while for chaotic orbits, $ S / \ln(r) $ keeps growing and it tends to the final value $ 1 $ as $ \gamma $ fills the cells of the partition.

The Shannon entropy thus serves as an efficient tool to quantify stability, but its greatest benefit is that the direct estimation of the diffusion coefficient of the chaotic diffusion is also feasible. Following \cite{cincotta2021febr} and \cite{cincotta2021march}, the local diffusion coefficient for the trajectory $ \gamma $ in the interval $ (t, t+\delta t) $ is given by
\begin{equation}
    D_S(t; \gamma) = \frac{(I_{\mathrm{max}} - I_{\mathrm{min}})^2}{r} r_0 \frac{\mathrm{d}S(\gamma, \alpha)}{\mathrm{d}t},
    \label{eq:localD}
\end{equation}
where $ I_\mathrm{max} $ and $ I_\mathrm{min} $ are the extrema of $ I(t) = \sqrt{I_1(t)^2 + I_2(t)^2} $, $ r_0 \equiv r_0(t) $ gives the number of non-empty cells in $ \alpha $, and $ \mathrm{d}S/\mathrm{d}t $ denotes the time derivative of the normalized entropy. The global diffusion coefficient of $ \gamma $ is then obtained by taking the time average of the local coefficient \eqref{eq:localD}:
\begin{equation}
	D_S(\gamma) = \lim_{t \to \infty} \frac{1}{t} \int_{t_0}^t D_S(t; \gamma) \mathrm{d}t \approx \langle D_S(t; \gamma)\rangle_{t\leq N}.
	\label{eq:globalD}
\end{equation}

In the knowledge of the global diffusion coefficient, the characteristic time of stability (or escape time) is approximated by the inverse of the latter:
\begin{equation}
	\tau_{\mathrm{esc}}(\gamma) = K\frac{\Delta^2}{D_S(\gamma)}.
	\label{eq:tau_esc}
\end{equation}
The proportionality factor in \eqref{eq:tau_esc} contains the mean square displacement $ \Delta^2 \equiv I_{1,\mathrm{lim}}^2 + I_{2,\mathrm{lim}}^2 $ (where $ I_{1,\mathrm{lim}} $ and $ I_{2,\mathrm{lim}} $ are the half-lengths of the sides of $ \alpha $), and the numerical constant $ K $ of order of unity in which factor the route of the diffusion in $ \mathcal{S} $ is taken into account.

We note here that the above derivation of the diffusion coefficient and stability time from the entropy requires the assumption that the diffusion is nearly normal, i.e., the time-dependence of the variance of any given phase variable is linear. The assumption is not always valid as was shown for example in \cite{cincotta2018}, but alas, their computation in the cases of the non-linear sub- and super-diffusion is yet an open and complicated question. However, locally (in a small neighbourhood of a given initial condition) and for relatively short timescales (the time required for reaching the steady state, i.e., a homogeneous and isotropic diffusion) the normal approximation is justified.

Let us, furthermore, remark that the quantities $ r $, $ I_{1,\mathrm{lim}} $, $ I_{2,\mathrm{lim}} $, and $ N $ are free parameters of the Shannon entropy method. \cite{cincotta2021febr} showed, however, that both
$ S $ and $ D_S $ (and hence $ \tau_\mathrm{esc} $, too) are nearly invariant with respect to the choice of the above parameters, whenever the inequality $ r_0 \ll N \lesssim r $ is satisfied, or equivalently, whenever $ N < r \lesssim N^{1/\hat{S}} $ holds (where $ \hat{S} < 1 $ is an empirically adopted threshold value of the normalized entropy).

\section{Results and discussion}
\label{sec:results_and_discussion}

\subsection{The computational setup}
\label{subsec:the_computational_setup}

For our computations, we used the orbital elements and physical parameters of the planets (and the latter of the central star) as given in \cite{gozdziewski2016}. The quantities are provided in Table~1 therein, separately for the case of the pure Laplace resonance and for the case of the chain of $ 2 $-body resonances. Since both models are coplanar with the inclinations $ i = 90^\circ $ and longitudes of ascending node $ \Omega = 0^\circ $ for all three of the planets, the general four-body problem is reduced to the planar 4BP, and only the action-like semi-major axes $ a $ and eccentricities $ e $, and the angle-like arguments of pericentre $ \omega $ and mean anomalies $ M $ remain as variables of the problem.

The dynamics of the two resonant configurations are being investigated in the forthcoming subsections - separately for the pure Laplace resonance in Section~\ref{subsec:the_pure_laplace_resonance} and for the case of the chain of $ 2 $-body resonances in Section~\ref{subsec:the_chain_of_2-body_resonances} - by means of $ 2 $-dimensional dynamical maps in the $ (a, e) $ plane. Grids of $ s = 100 \times 100 $ initial conditions (ICs) were constructed around the nominal positions $ (a_i, e_i) $ ($ i = b,c,d $) of the planets \citep[see Table~1 in][]{gozdziewski2016}, with the boundaries $ a_i \pm 0.0025 $~AU ($ i = b,c,d $) in the direction of $ a $ and $ [0, 0.08] $ in the direction of $ e $. Each IC was integrated up to $ T_\mathrm{tot} = 3.2 \cdot 10^4 $~yrs (which is the same as in \citeauthor{gozdziewski2016} \citeyear{gozdziewski2016} and corresponds to $ 10^6 $ orbital periods of the outermost planet). We introduce here the notation 'regular' for those initial condition points that reached the end of integration, to distinguish them from those that became unstable earlier. Those ICs that are not 'regular' in this sense, are depicted with grey in the figures. (During the integration of the ICs, we varied only the actions $ a $ and $ e $, and kept the angles $ \omega $ and $ M $ as well as the physical parameters of the two models of \citeauthor{gozdziewski2016} \citeyear{gozdziewski2016} unchanged.) The computations were carried out by the MERCURY $ n $-body integrator \citep{chambers1999}, using its Bulirsch--Stoer routine with a sampling timestep of $ h = 0.1 $~yrs and a precision of $ 10^{-12} $.

The grids of initial conditions having been constructed, the computation of the Shannon entropy for a given IC at the position $ (a_0, e_0) $ was implemented as follows. The actions $ I_1 $ and $ I_2 $ (see Section~\ref{sec:the_shannon_entropy_formalism}) were set to be the semi-major axis $ a $ and eccentricity $ e $, whereas the argument of pericentre $ \omega $ and mean anomaly $ M $ represent now the phases $ \vartheta_1 $ and $ \vartheta_2 $. By keeping the latter fixed at their nominal values \citep[][Table 1]{gozdziewski2016}, the section $ \mathcal{S} $ is defined. The partition $ \alpha $ around $ (a_0, e_0) $ was composed as $ a_0 \pm a_\mathrm{lim} $, $ e_0 \pm e_\mathrm{lim} $\footnote{In case $ e_0-e_\mathrm{lim} $ resulted in a negative value, we adopted the borders $ [0, 2 e_\mathrm{lim}] $ instead of the originals.} with the boundaries
\begin{align}
    a_\mathrm{lim} &= 2\sqrt{3} R_\mathrm{H},
    \label{eq:alim}\\
    e_\mathrm{lim} &= \frac{1}{2} \max_{i=1, \dots,s}\{\Delta e_i\text{: the IC is 'regular'}\}.
    \label{eq:elim}
\end{align}
In \eqref{eq:alim}, $ R_\mathrm{H} $ denotes the mutual Hill radius of two adjacent planets (of masses $ m_i $ $ (i=1,2) $ and semi-major axes $ a_i $ $ (i=1,2) $) defined as
\begin{equation}
    R_\mathrm{H} = \left(\frac{m_1+m_2}{3m_*}\right)^{1/3}\frac{a_1+a_2}{2}
    \label{eq:RHill}
\end{equation}
($ m_* $ is the stellar mass), and
\begin{equation}
    \Delta e_i = \max\limits_{t\leq T_\mathrm{tot}}(e_i)-\min\limits_{t\leq T_\mathrm{tot}}(e_i), \quad i = 1,\dots,s
    \label{eq:eccvar}
\end{equation}
in \eqref{eq:elim} represents the eccentricity variations of the initial condition points during the total time span of the integration, hence the expression $ \{\Delta e_i\text{: the IC is 'regular'}\} $ selects the eccentricity variations of the 'regular' points - in the sense as introduced above. The number of cells of the so-defined partition was set to $ r= 800\times800 $.

As already noted in Section~\ref{sec:the_shannon_entropy_formalism}, $ r $, $ a_\mathrm{lim} $, $ e_\mathrm{lim} $, and $ N $ are free parameters of the entropy method. Our choices of them were such that the required inequality $ r_0(t\leq T_\mathrm{tot})\ll N \lesssim r $ was fulfilled via $ \sim 4000 \ll 3.2\cdot10^5 \lesssim 6.4\cdot10^5 $ (where $ 4000 $ is a mean value of $ r_0 $ at $ T_\mathrm{tot} $, and the number of intersections $ N =3.2\cdot10^5 $ equals, in practice, to the ratio of the total integration time $ T_\mathrm{tot} $ and the sampling timestep $ h $). That is, first we set $ T_\mathrm{tot} $ to the desired value of $ 3.2\cdot10^4 $~yrs (see above), then chose $ h = 0.1 $~yrs so that their ratio was sufficiently large yet computationally feasible when dealing with tens of thousands of ICs. Doing so gave a lower boundary $ r_l \sim 600\times 600 $ for the number of elements of the partitions \citep[and as for the upper limit $ r_u \sim 5000\times 5000 $, it was obtained from the second inequality of][see Section~\ref{sec:the_shannon_entropy_formalism}]{cincotta2021febr}. The boundaries of the partitions were chosen according to physical considerations (see Equations \eqref{eq:alim}--\eqref{eq:eccvar}).

In regard to the numerical computation of the diffusion coefficients and the escape times, we approximated the derivative of the entropy in \eqref{eq:localD} with the second-order central (or symmetric) difference quotient, and as for the constant $ K $ in \eqref{eq:tau_esc}, we equated it with $ 1 $.

\subsection{The pure Laplace resonance}
\label{subsec:the_pure_laplace_resonance}

\subsubsection{The phase space of planet~$ b $}
\label{subsubsec:the_phase_space_of_planet_b}

\begin{figure*}
    \centering
    \begin{subfigure}{0.33\textwidth}
		\includegraphics[width = \textwidth]{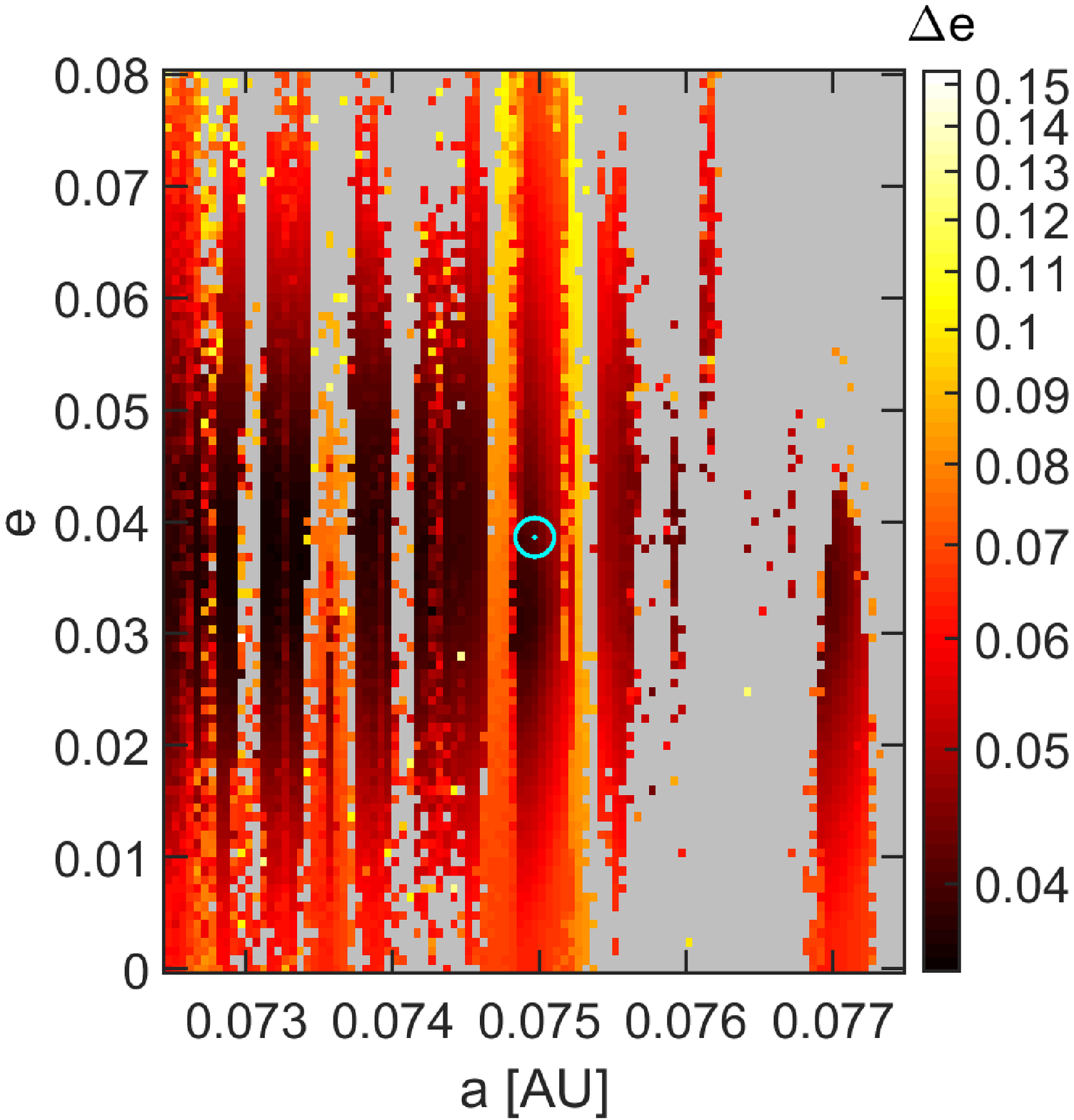}
	\end{subfigure}
	\qquad
	\begin{subfigure}{0.33\textwidth}
		\includegraphics[width = \textwidth]{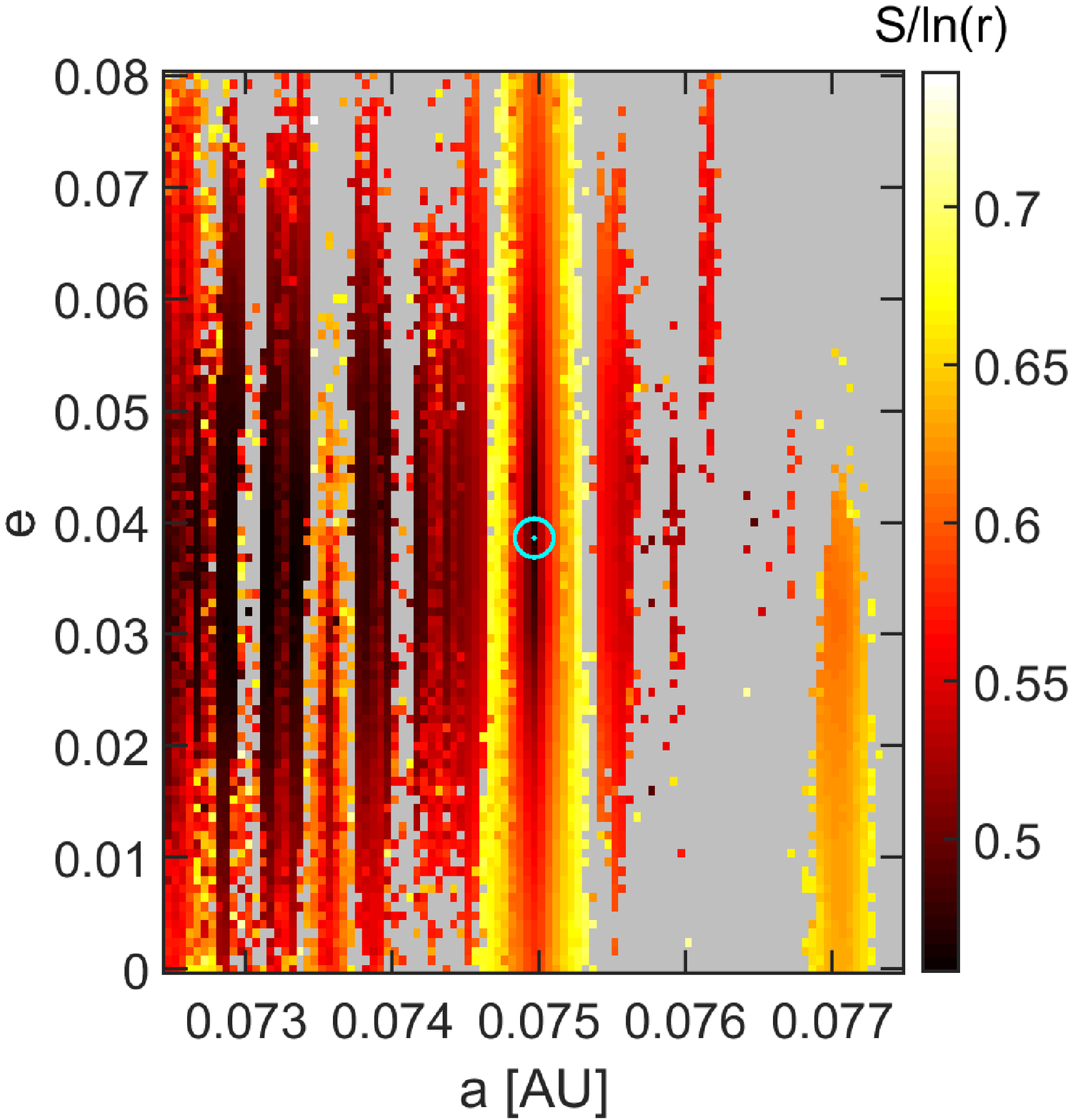}
	\end{subfigure}
	\qquad
	\begin{subfigure}{0.33\textwidth}
	    \includegraphics[width = \textwidth]{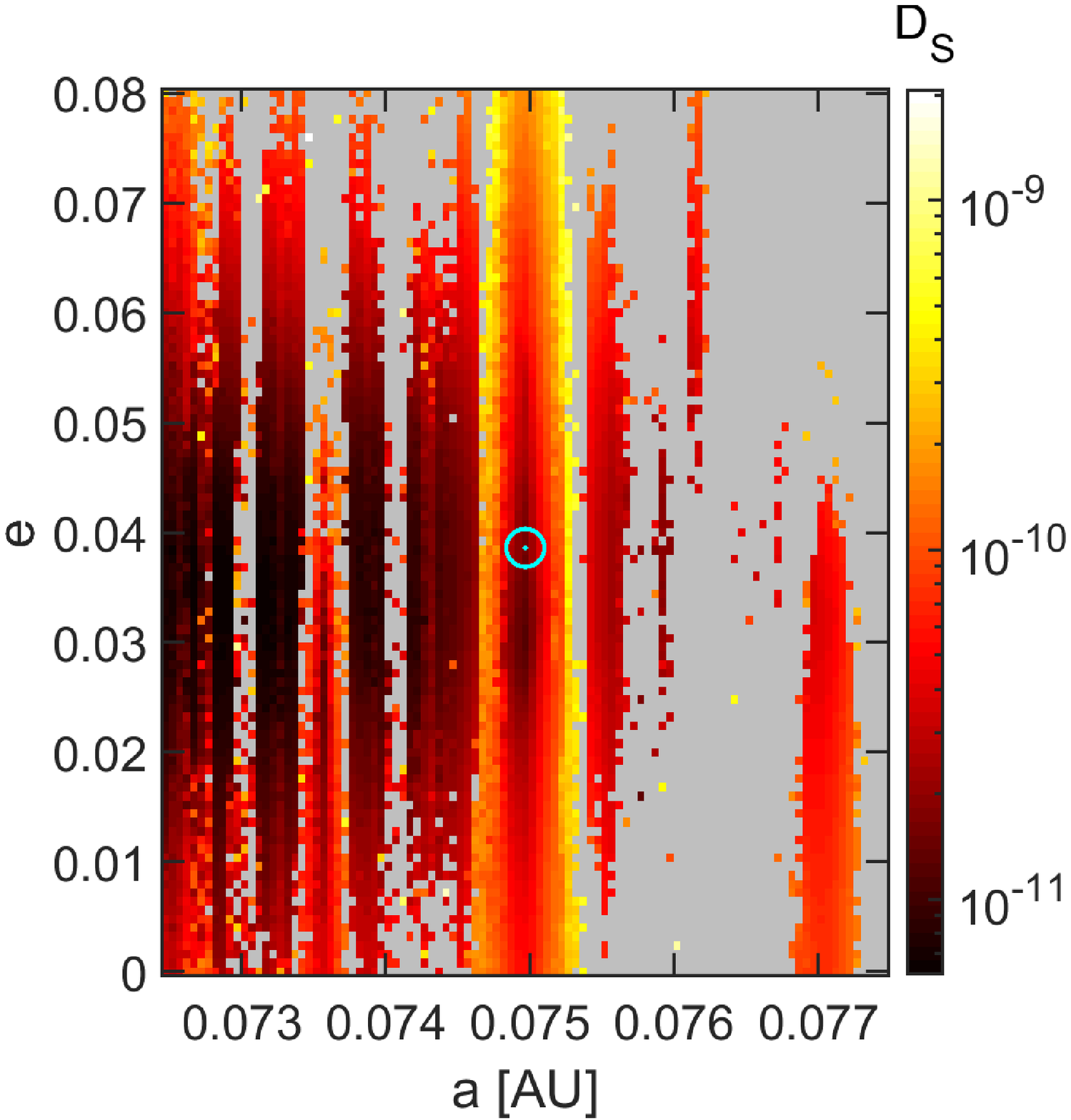}
	\end{subfigure}
	\qquad
	\begin{subfigure}{0.33\textwidth}
		\includegraphics[width = \textwidth]{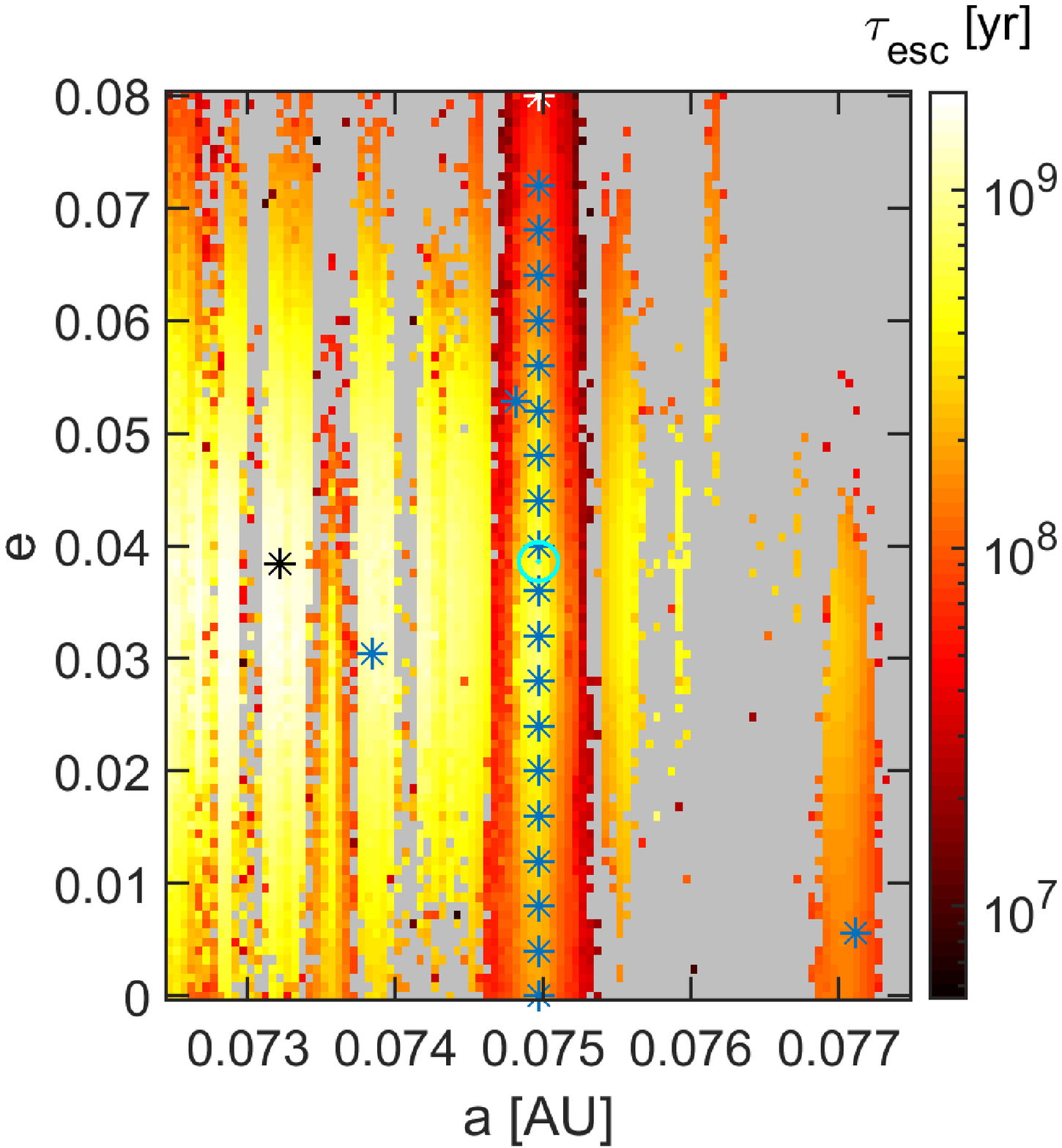}
	\end{subfigure}
	\caption{Dynamical maps for Kepler-60$b$, assuming a pure Laplace resonance. The nominal position of planet~$ b $ is denoted with cyan-coloured circles at $ a_{b1} = 0.07497 $~AU, $ e_{b1} = 0.0386 $. The grey-coloured points did not reach the end of integration, i.e., with our previous denomination, they are not 'regular'. \textbf{Upper left panel:} eccentricity variations throughout the integration. \textbf{Upper right panel:} normalized Shannon entropy at the end of integration. \textbf{Lower left panel:} the global diffusion coefficients of the chaotic diffusion. \textbf{Lower right panel:} stability times derived from the diffusion coefficients. (The asterisk symbols mark the initial conditions of long-term, direct integrations (see Section~\ref{subsec:comparison_with_direct_long_term_integrations}).)}
    \label{fig:K60B1}
\end{figure*}

In the framework of the pure Laplacian resonant model, we begin our discussion with planet~$ b $, the innermost body of the system.

The results are shown in Figure~\ref{fig:K60B1}. The upper left panel is the map of the eccentricity variations $ \Delta e $ (see Equation~\eqref{eq:eccvar}). This quantity serves as a reliable and widely used indicator of stability \citep[see e.g.][]{marti2013,beauge2019}, since chaos is almost always accompanied by large excursions in the eccentricity. Thus we used this simple check to have an a priori knowledge of the structure of the phase space. The non-grey 'regular' points appear to be particularly stable. Even the largest values of $ \Delta e $ remain below $ \sim 0.15 $, but in the most stable regions of the figure, values less than $ \sim 0.05 $ are not rare either. One observes that the islands of stability in the figure turn up as vertical stripes stretching from $ e = 0 $ to $ 0.08 $. The widest among them is associated with the Laplace MMR around the nominal position of planet~$ b $ at $ a_{b1} = 0.07497 $~AU, $ e_{b1} = 0.0386 $ (see the cyan circle in the figure), and the others are probable manifestations of higher-order mean-motion resonances or secondary, possibly secular resonances.

The three other panels of Figure~\ref{fig:K60B1} display quantities derived from the Shannon entropy. The normalized entropy $ S/\ln(r) $ is shown in the upper right panel. The values displayed in the figure are those reached at the end of the total integration time, i.e., $ S(t=T_\mathrm{tot})/\ln(r) $ is being shown. As anticipated by having studied the map of $ \Delta e $, the Shannon entropy reveals similarly stable dynamics in the 'regular' domains of the phase space. In the vast majority of the panel, the entropy values remain below $ \sim 0.55 $. Yet, there is an important difference in comparison to the previous map that should be pinpointed here. Whilst the centre of the main resonance was only moderately appointed by the eccentricity variations, the entropy values pointedly emphasize it. This extra information received here is related to the Shannon entropy being particularly sensitive to resonances.

The left panel in the bottom row of Figure~\ref{fig:K60B1} shows the global diffusion coefficients $ D_S $ (see Equation~\eqref{eq:globalD}), and the right one the escape times $ \tau_{\mathrm{esc}} $ (see Equation~\eqref{eq:tau_esc}). Both panels exhibit values in accordance with those in the previous two maps, indicating the stabilizing role of the Laplace resonance. The escape times within the dominant island in the last panel reach values as high as a few times $ 10^{8} $~yrs, but we encounter equally high (or even higher) values of $ \tau_\mathrm{esc} $ in certain regions outside of the primary resonance, too. The shortest stability times among the non-grey 'regular' points are $ \sim 10^6 - 10^7 $~yrs. (Remark: since $ \tau_\mathrm{esc} $ is the reciprocal of $ D_S $ (except for a constant factor), these two bottom panels are, in fact, the same but with inverse colouring and different scaling. On account of this, in the subsequent figures we omit the map of the diffusion coefficients and present only the three other panels.)

\begin{figure*}
    \centering
    \begin{subfigure}{0.33\textwidth}
		\includegraphics[width = \textwidth]{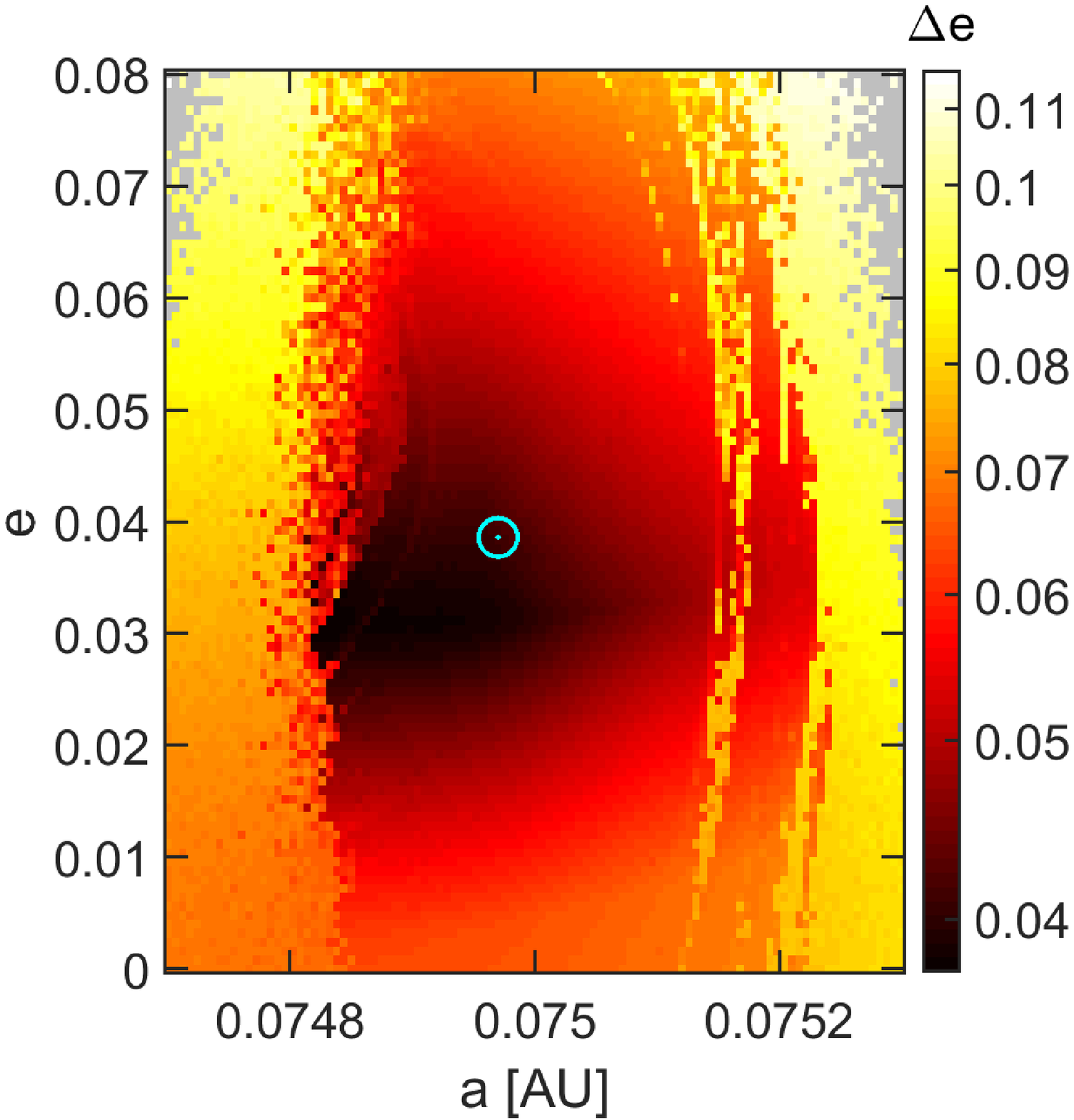}
	\end{subfigure}
	\begin{subfigure}{0.33\textwidth}
		\includegraphics[width = \textwidth]{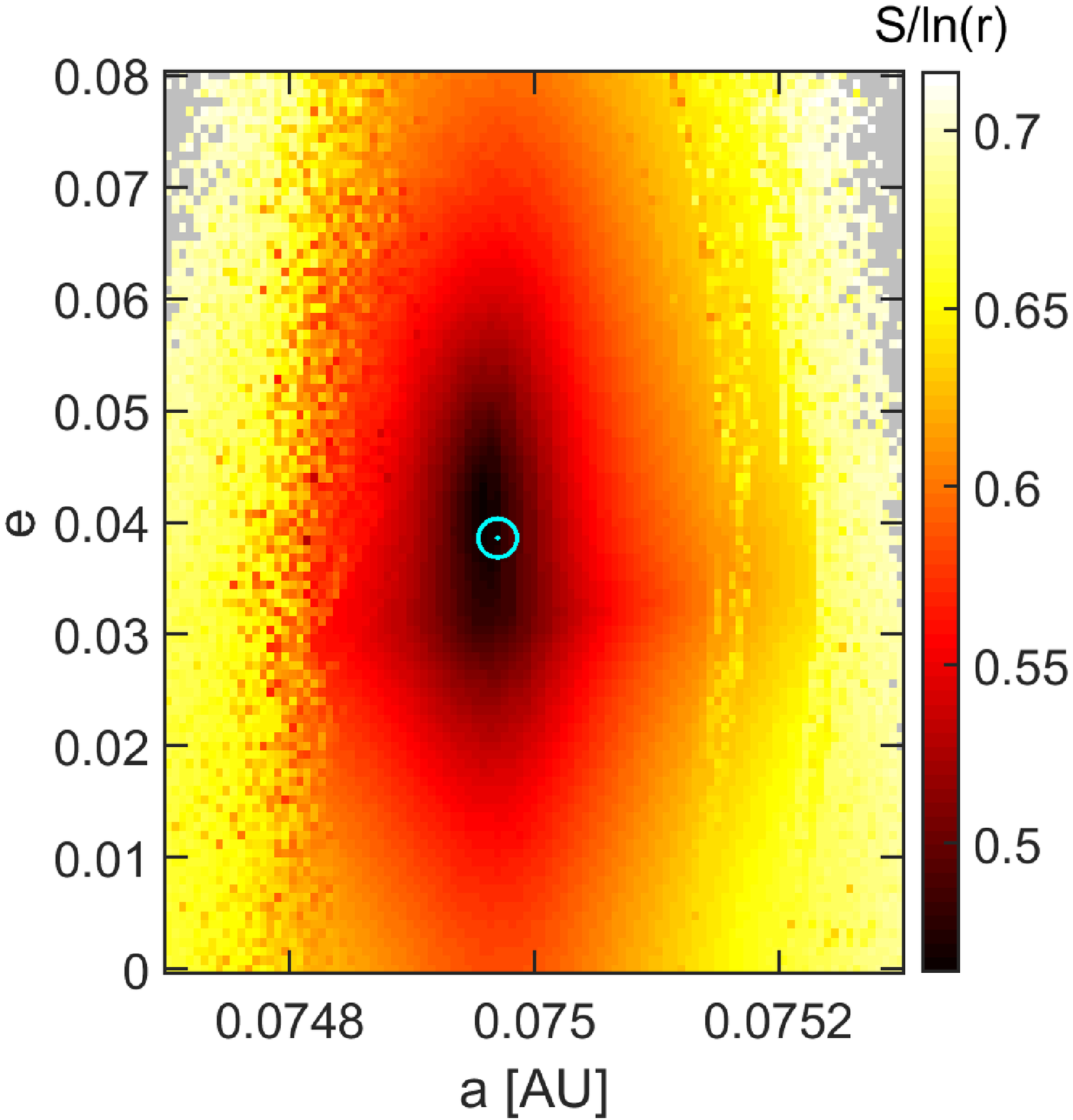}
	\end{subfigure}
	\begin{subfigure}{0.33\textwidth}
	    \includegraphics[width = \textwidth]{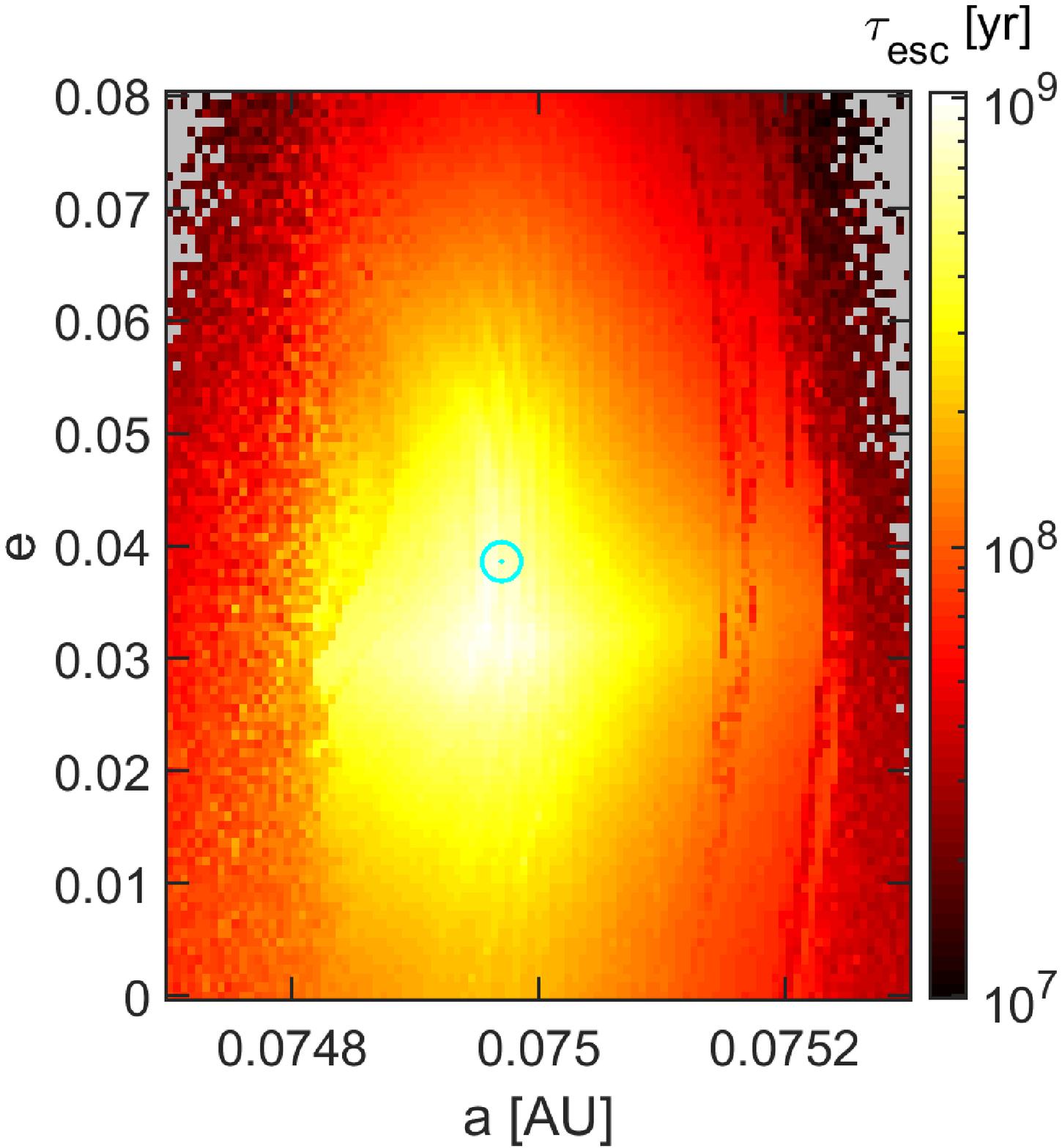}
	\end{subfigure}
	\caption{A magnification of Figure~\ref{fig:K60B1} around the centre of the pure Laplace resonance. \textbf{Left panel:} eccentricity variations throughout the integration. \textbf{Middle panel:} normalized Shannon entropy at the end of integration. \textbf{Right panel:} stability times derived from the diffusion coefficients.}
    \label{fig:K60B1narrow}
\end{figure*}

Before proceeding to discuss our results of the middle planet, we present a magnification of the resonant centre around planet~$ b $ in Figure~\ref{fig:K60B1narrow} consisting now of only three panels: the maps of $ \Delta e $, $ S/\ln(r) $, and $ \tau_\mathrm{esc} $. The boundaries of the grid of initial conditions in this set of panels are $ [0.0747, 0.0753] $~AU in $ a $ and (unaltered) $ [0, 0.08] $ in $ e $. All the other parameters of the computations are as listed in Section~\ref{subsec:the_computational_setup}. The motivation of the magnification - apart from allowing a closer examination of the central part of the $ 3 $-body resonance - is to link our computations to those of \cite{gozdziewski2016}, in particular, to their second panel of Figure 4, where the authors investigated the same segment of the $ (a, e) $ plane by using the MEGNO chaos indicator. The similarity in the general structure of our figure and that of \cite{gozdziewski2016} is apparent. Some vertical arcs, that remained hidden in Figure~\ref{fig:K60B1}, are revealed here, one of them being particularly noticeable at $ \sim 0.07515 $~AU. The depicted values on the maps are in accordance with those of Figure~\ref{fig:K60B1}.

\subsubsection{The phase space of planet~$ c $}
\label{subsubsec:the_phase_space_of_planet_c}

\begin{figure*}
    \begin{subfigure}{0.33\textwidth}
		\includegraphics[width = \textwidth]{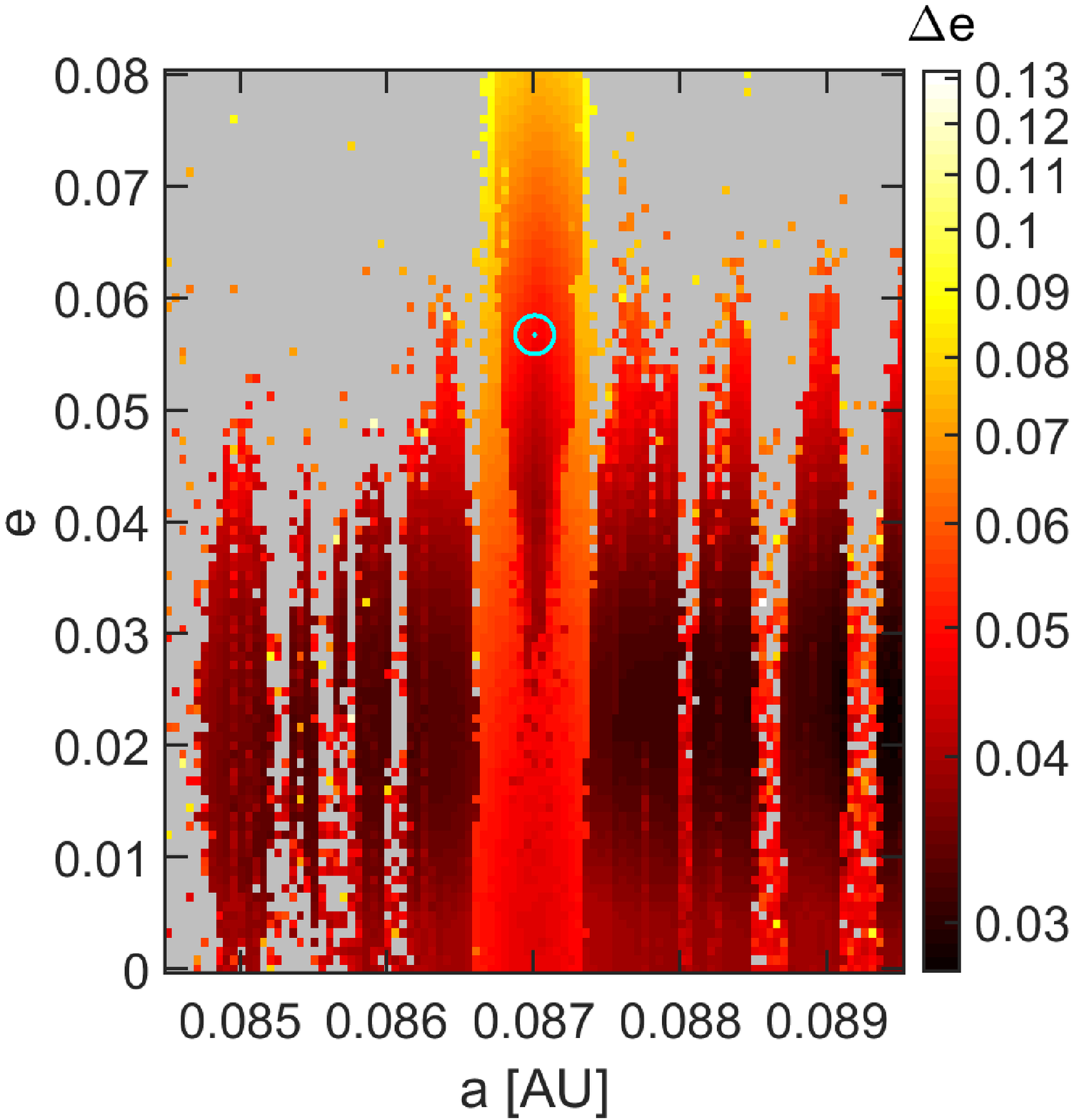}
	\end{subfigure}
	\begin{subfigure}{0.33\textwidth}
		\includegraphics[width = \textwidth]{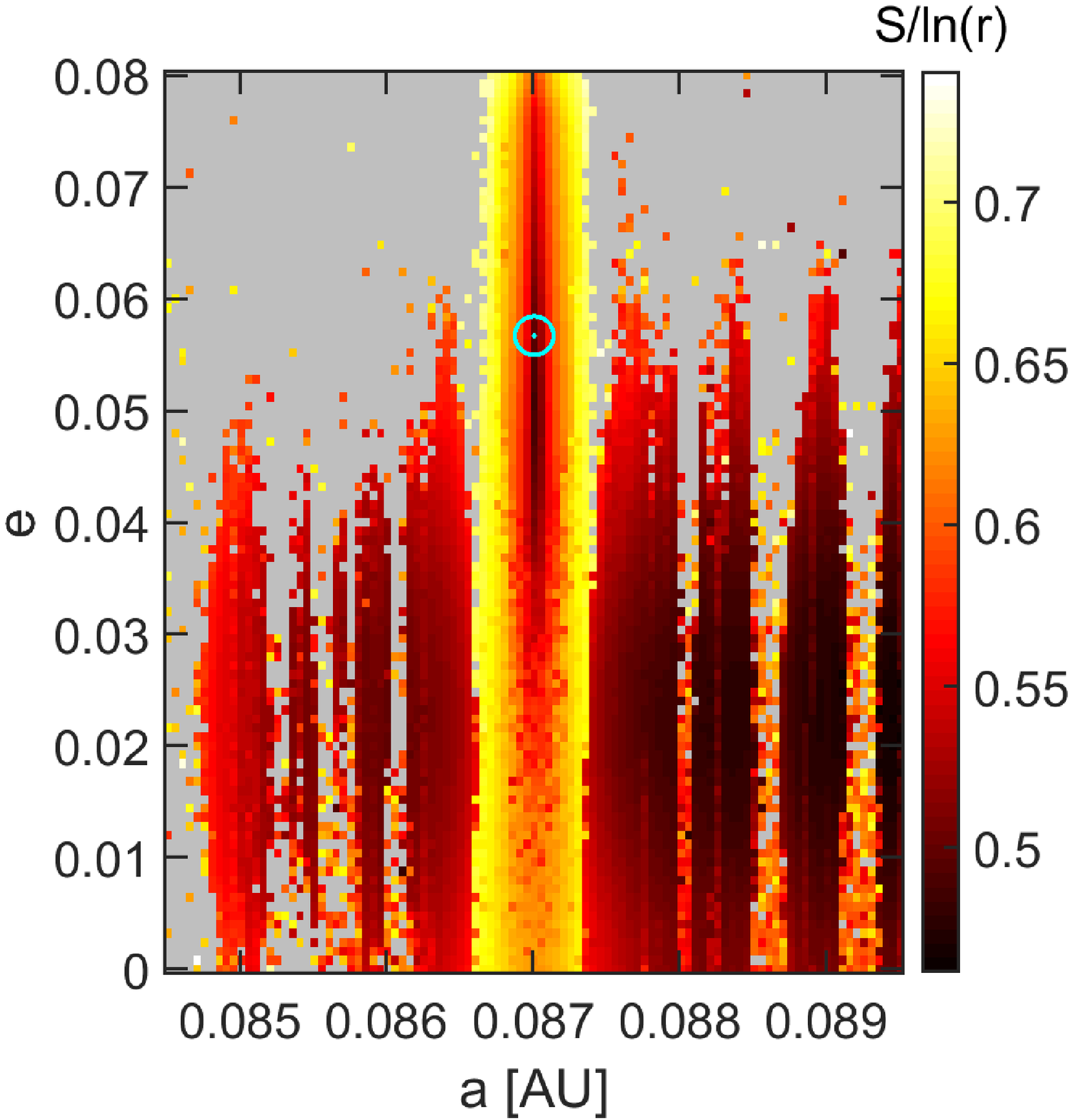}
	\end{subfigure}
	\begin{subfigure}{0.33\textwidth}
	    \includegraphics[width = \textwidth]{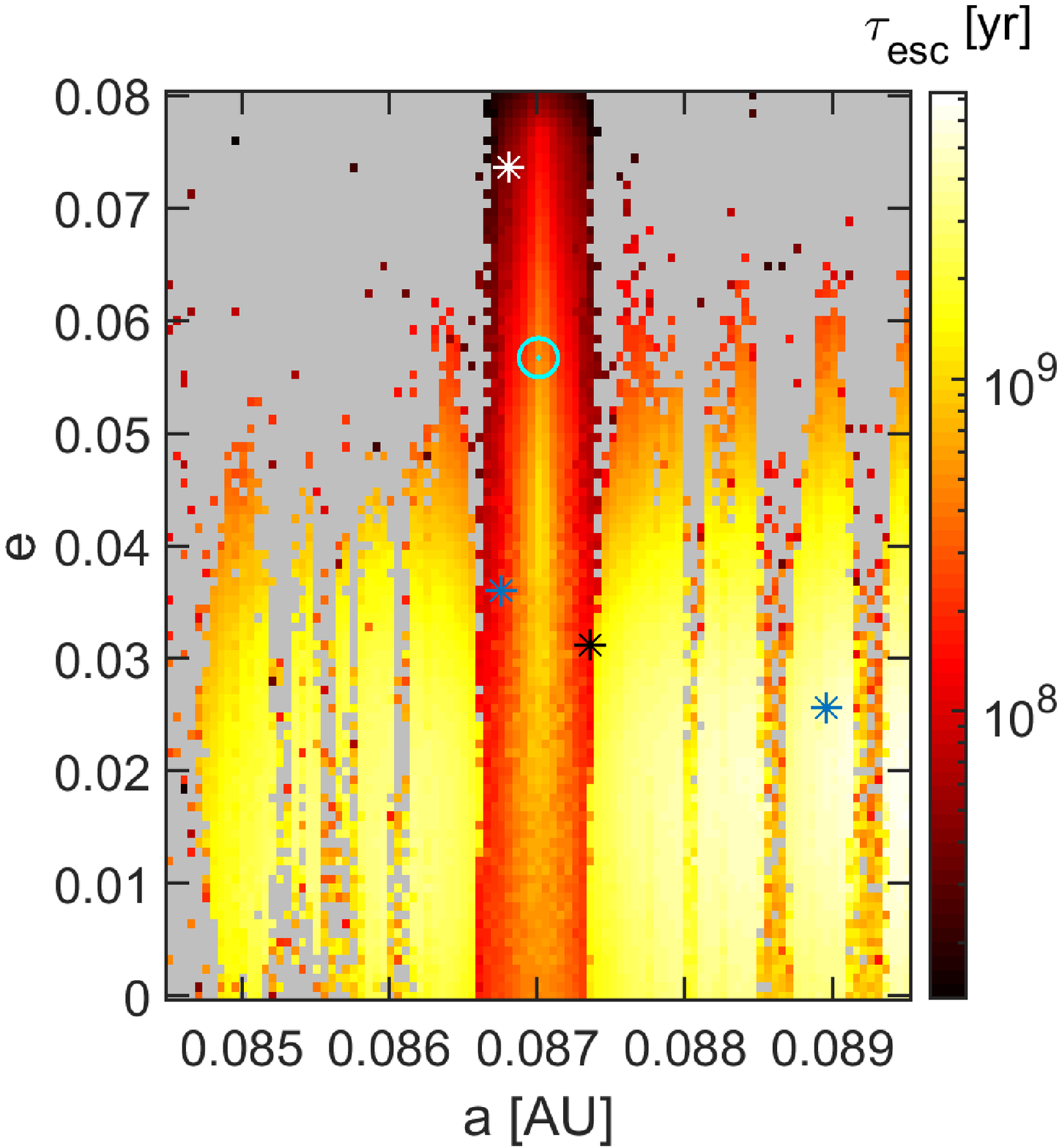}
	\end{subfigure}
	\caption{Dynamical maps for Kepler-60$c$, assuming a pure Laplace resonance. The nominal position of planet~$ c $ is denoted with cyan-coloured circles at $ a_{c1} = 0.08701 $~AU, $ e_{c1} = 0.0567 $. The grey-coloured points did not reach the end of integration, i.e., with our previous denomination, they are not 'regular'. \textbf{Left panel:} eccentricity variations throughout the integration. \textbf{Middle panel:} normalized Shannon entropy at the end of integration. \textbf{Right panel:} stability times derived from the diffusion coefficients. (The asterisk symbols mark the initial conditions of long-term, direct integrations (see Section~\ref{subsec:comparison_with_direct_long_term_integrations}).)}
    \label{fig:K60C1}
\end{figure*}

Although the dynamics of the three planets are not independent and thus the results of planet~$ b $ implicitly contain those of planets~$ c $ and $ d $ (all the more so that we stopped the integrations whenever a close encounter took place, regardless of the pair concerned), it is worth inspecting the phase space in the neighbourhood of the outer planets, too. Hence, in the present subsection, we discuss our results regarding the middle body of the system.

Figure~\ref{fig:K60C1} shows the dynamical maps for planet~$ c $, assuming a pure Laplace resonance. The general appearance of this part of the phase space is slightly different from the one in the vicinity of planet~$ b $. Although vertically elongated stable zones are likewise present, they discontinue at eccentricities $ \sim0.05 $, save the primary island of the Laplace resonance in the middle, enclosing planet~$ c $ at the nominal position $ a_{c1} = 0.08701 $~AU, $ e_{c1} = 0.0567 $ (see the cyan-coloured circles in the figure). The latter region is characterized by very similar values to those measured in the case of planet~$ b $: close to the centre of the resonance the eccentricity variations $ \Delta e $ are below $ \sim 0.05 - 0.06 $, the normalized Shannon entropy $ S/\ln(r) $ at the end of the integration is no larger than $ \sim 0.5 - 0.55 $, and the stability times $ \tau_\mathrm{esc} $ yield a few times $ 10^8 $~yrs. And again, in some low-eccentricity regions outside the primary $ 3 $-body resonance, these values are exceeded and the escape times reach even $ \sim 10^{9} $~yrs.

\subsubsection{The phase space of planet~$ d $}
\label{subsubsec:the_phase_space_of_planet_d}

\begin{figure*}
    \begin{subfigure}{0.33\textwidth}
		\includegraphics[width = \textwidth]{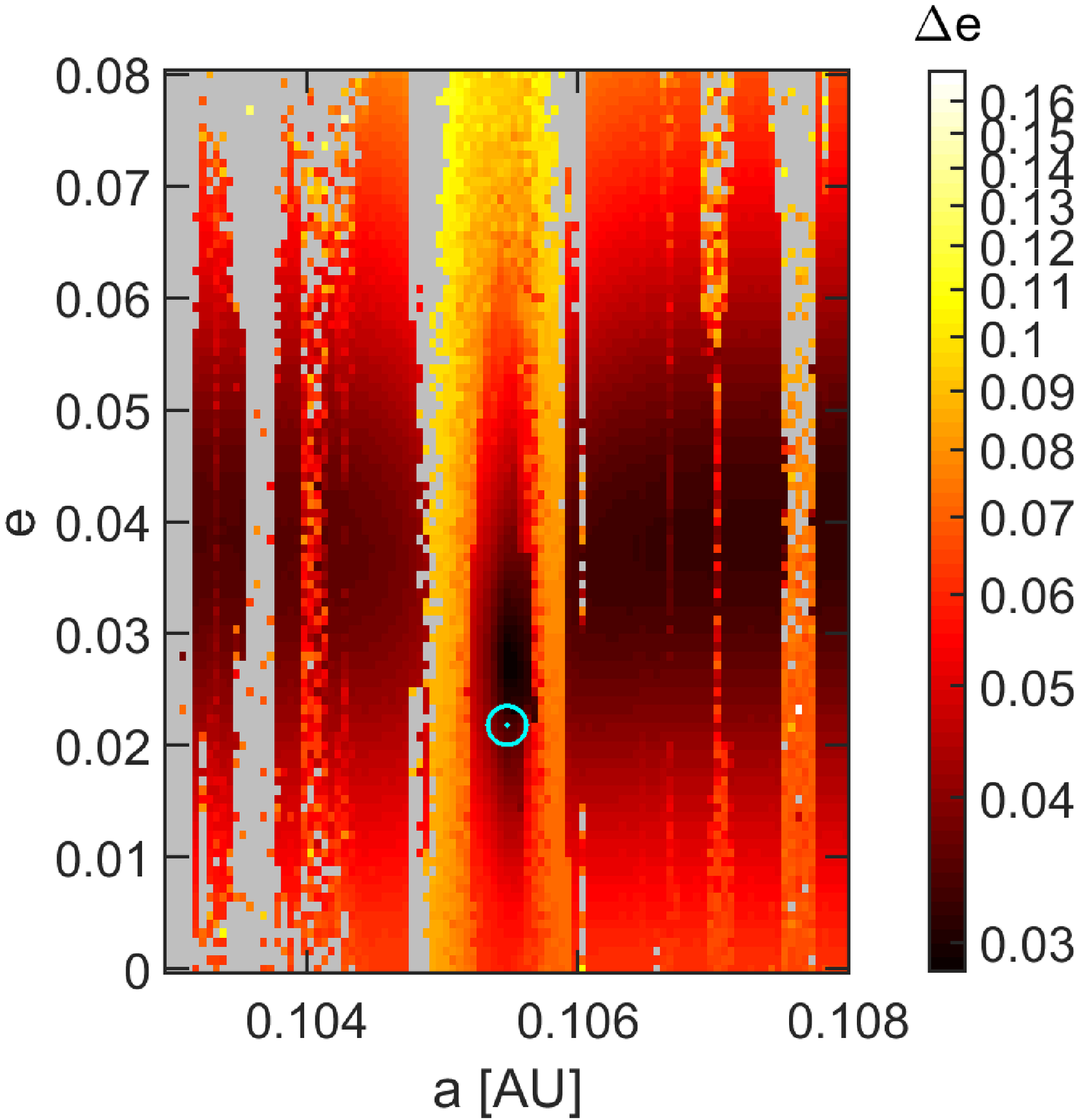}
	\end{subfigure}
	\begin{subfigure}{0.33\textwidth}
		\includegraphics[width = \textwidth]{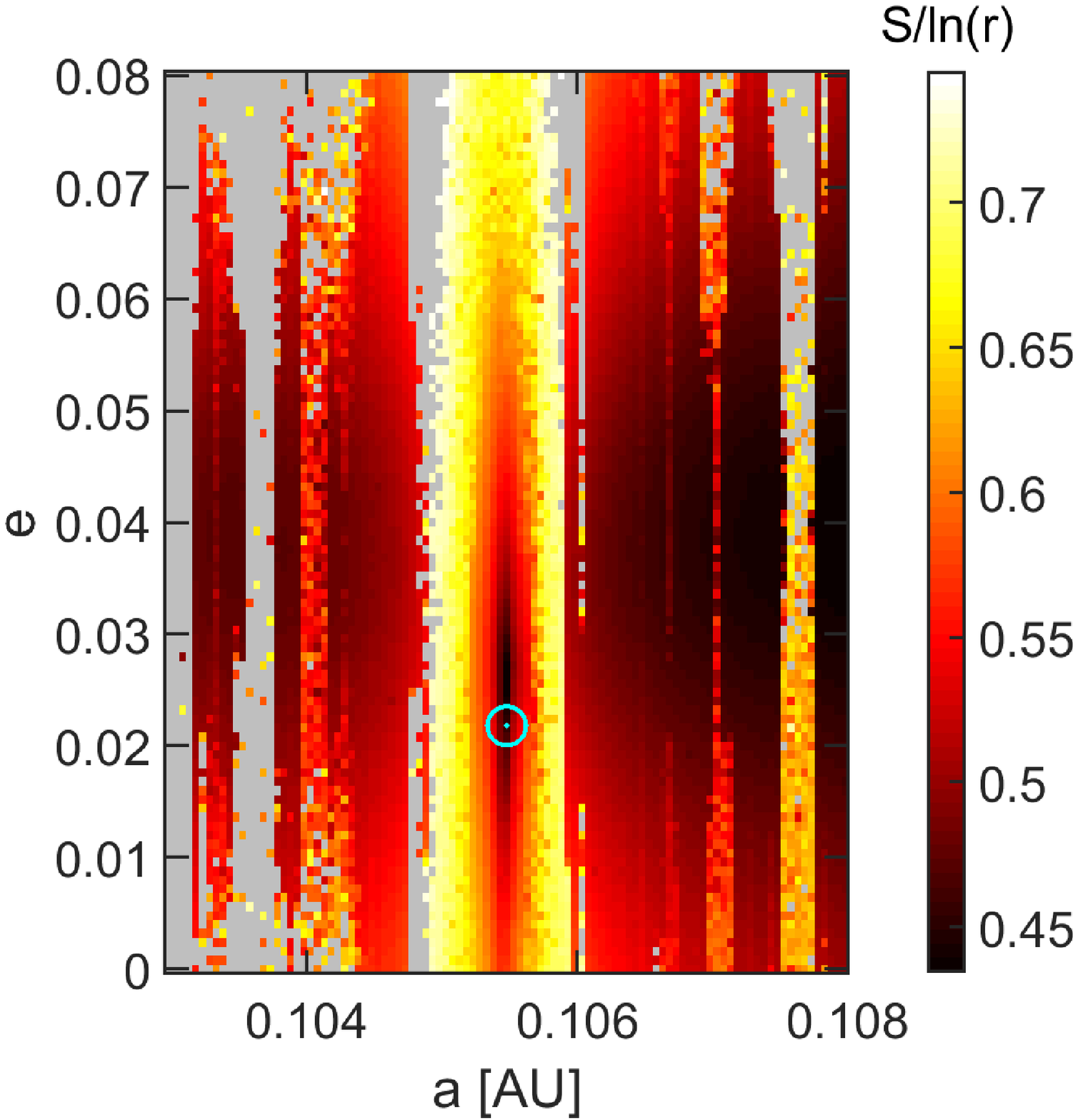}
	\end{subfigure}
	\begin{subfigure}{0.33\textwidth}
		\includegraphics[width = \textwidth]{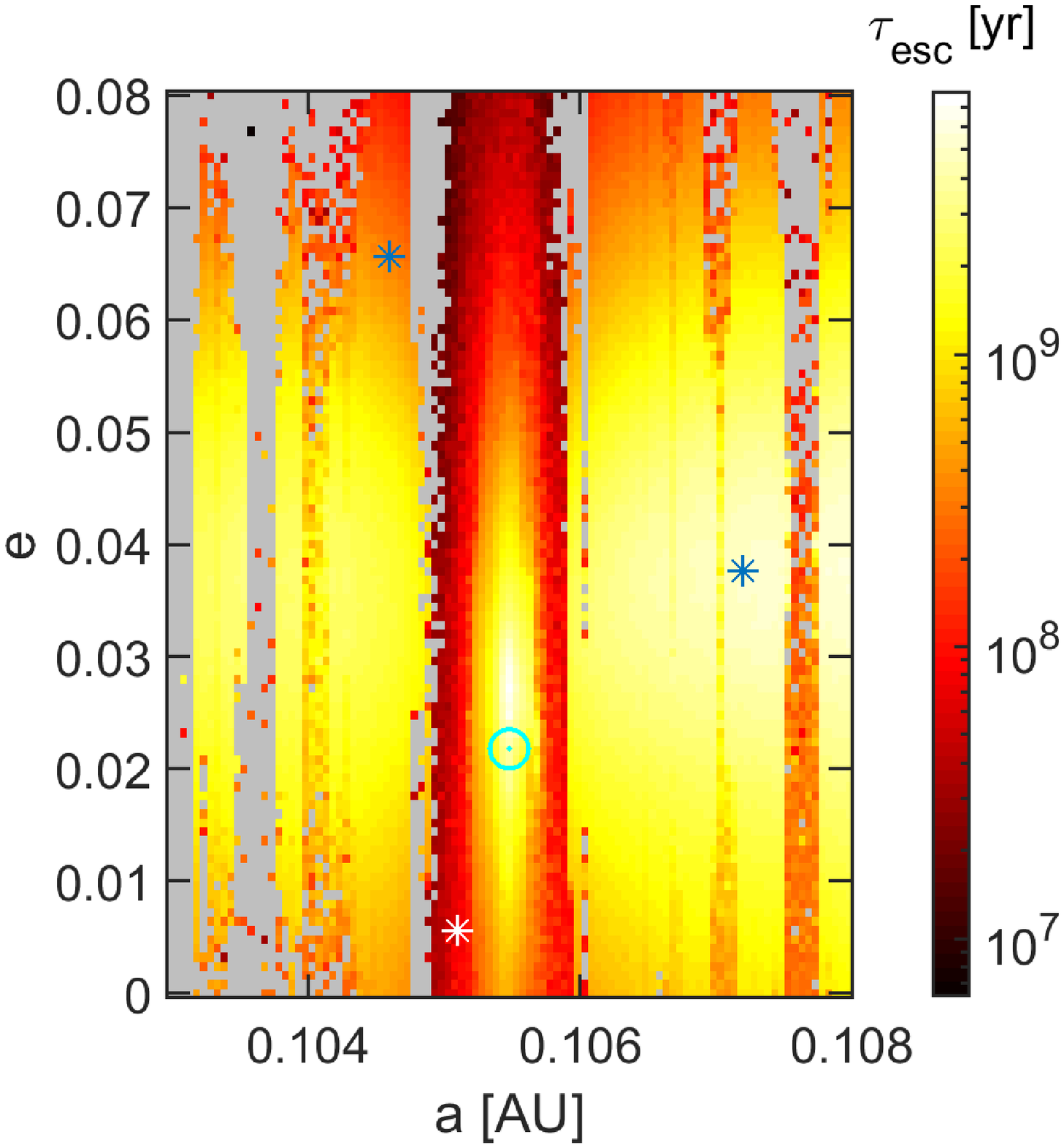}
	\end{subfigure}
	\caption{Dynamical maps for Kepler-60$d$, assuming a pure Laplace resonance. The nominal position of planet~$ d $ is denoted with cyan-coloured circles at $ a_{d1} = 0.10548 $~AU, $ e_{d1} = 0.0218 $. The grey-coloured points did not reach the end of integration, i.e., with our previous denomination, they are not 'regular'. \textbf{Left panel:} eccentricity variations throughout the integration. \textbf{Middle panel:} normalized Shannon entropy at the end of integration. \textbf{Right panel:} stability times derived from the diffusion coefficients. (The asterisk symbols mark the initial conditions of long-term, direct integrations (see Section~\ref{subsec:comparison_with_direct_long_term_integrations}).)}
    \label{fig:K60D1}
\end{figure*}

The results for the outermost planet are shown in Figure~\ref{fig:K60D1}. The nominal position of planet~$ d $ in the case of the pure Laplace resonance is $ a_{d1} = 0.10548 $~AU, $ e_{d1} = 0.0218 $ (see the cyan-coloured circles in the figure). The three panels show a, by now familiar, stripe-patterned phase space segment with almost vertical, and this time somewhat broader, bands of stable motion stretching through the whole length of the maps. This observation indicates the presence of several higher-order, secondary, or secular resonances that govern the dynamics of the outer regions of the system. The non-'regular', grey-coloured points are now a little less numerous than in the cases of the two inner planets, and as for the 'regular' ones, we note values of the same order as before. The centre of the pure Laplace resonance is characterized by $ \Delta e \lesssim 0.04 $, $ S/\ln(r) \sim 0.45-0.5 $, $ \tau_{\mathrm{esc}} \sim 10^{9} $~yrs.

\subsection{The chain of 2-body resonances}
\label{subsec:the_chain_of_2-body_resonances}

After discussing the case of the pure Laplace resonance, we start analyzing our results related to the chain of two $ 2 $-body resonances.

The orbital elements and physical parameters of the planets were obtained again from Table 1 of \cite{gozdziewski2016}. Here one observes that the most important difference between the two fits of the authors is that in the case of the two $ 2 $-body resonances the eccentricities of the planets are slightly lower than in the case of the pure Laplace resonance. This finding might already suggest a more stable phase space but not necessarily.

The computations - similarly to the previous case - were performed separately for the three planets. We start reviewing the results of planet~$ b $.

\subsubsection{The phase space of planet~$ b $}
\label{subsubsec:the_phase_space_of_planet_b2}

\begin{figure*}
    \begin{subfigure}{0.33\textwidth}
		\includegraphics[width = \textwidth]{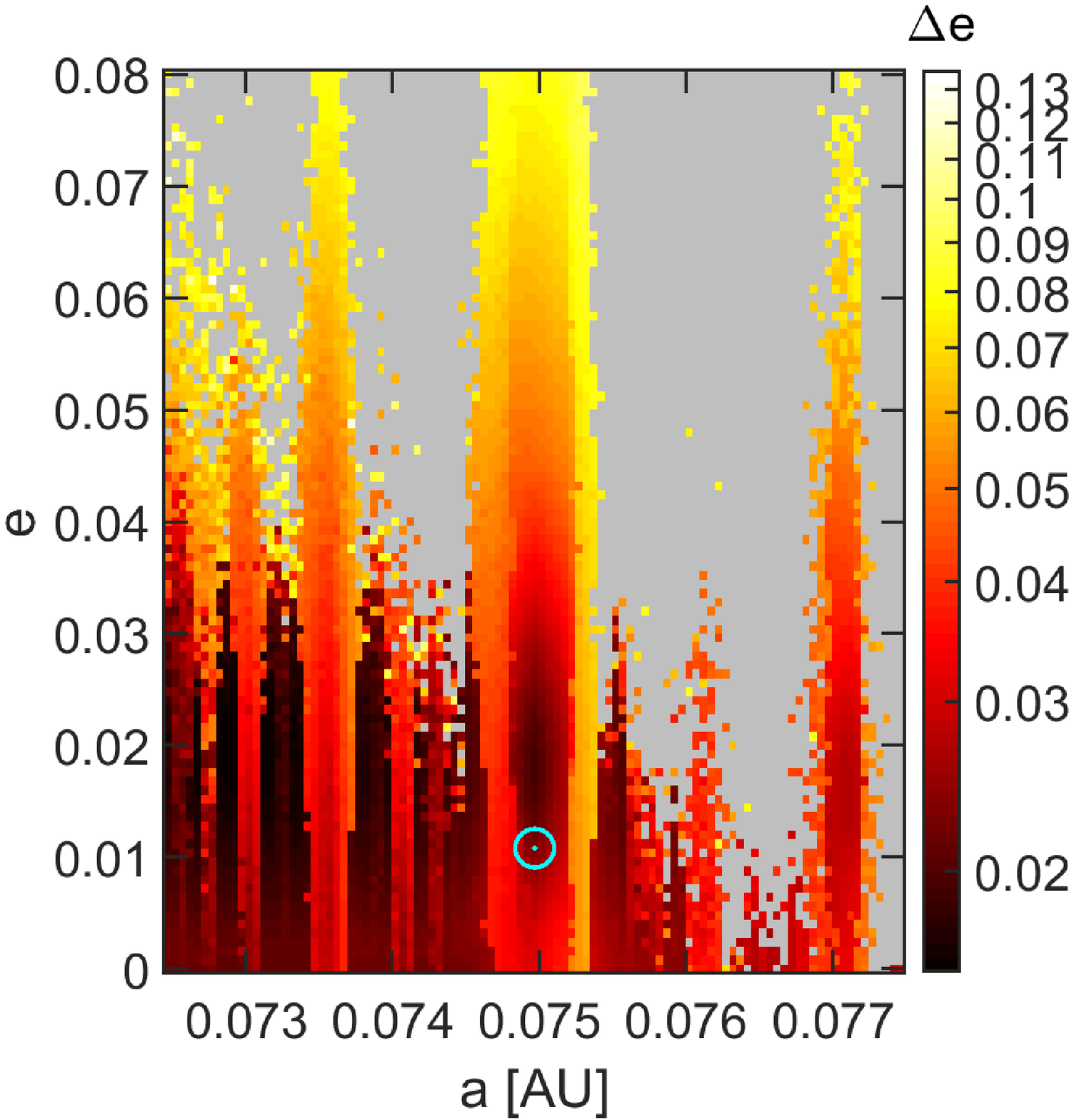}
	\end{subfigure}
	\begin{subfigure}{0.33\textwidth}
		\includegraphics[width = \textwidth]{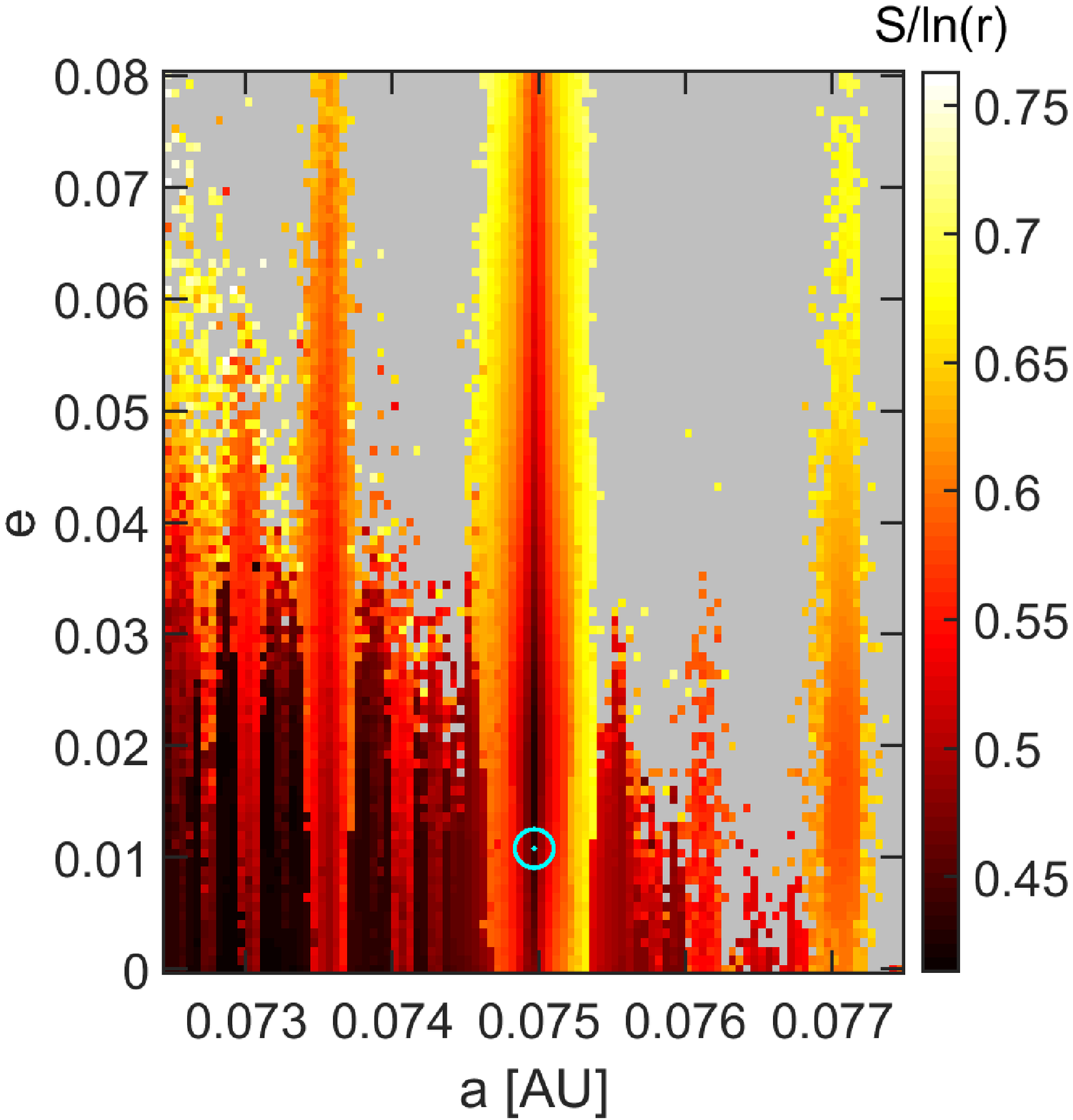}
	\end{subfigure}
	\begin{subfigure}{0.33\textwidth}
		\includegraphics[width = \textwidth]{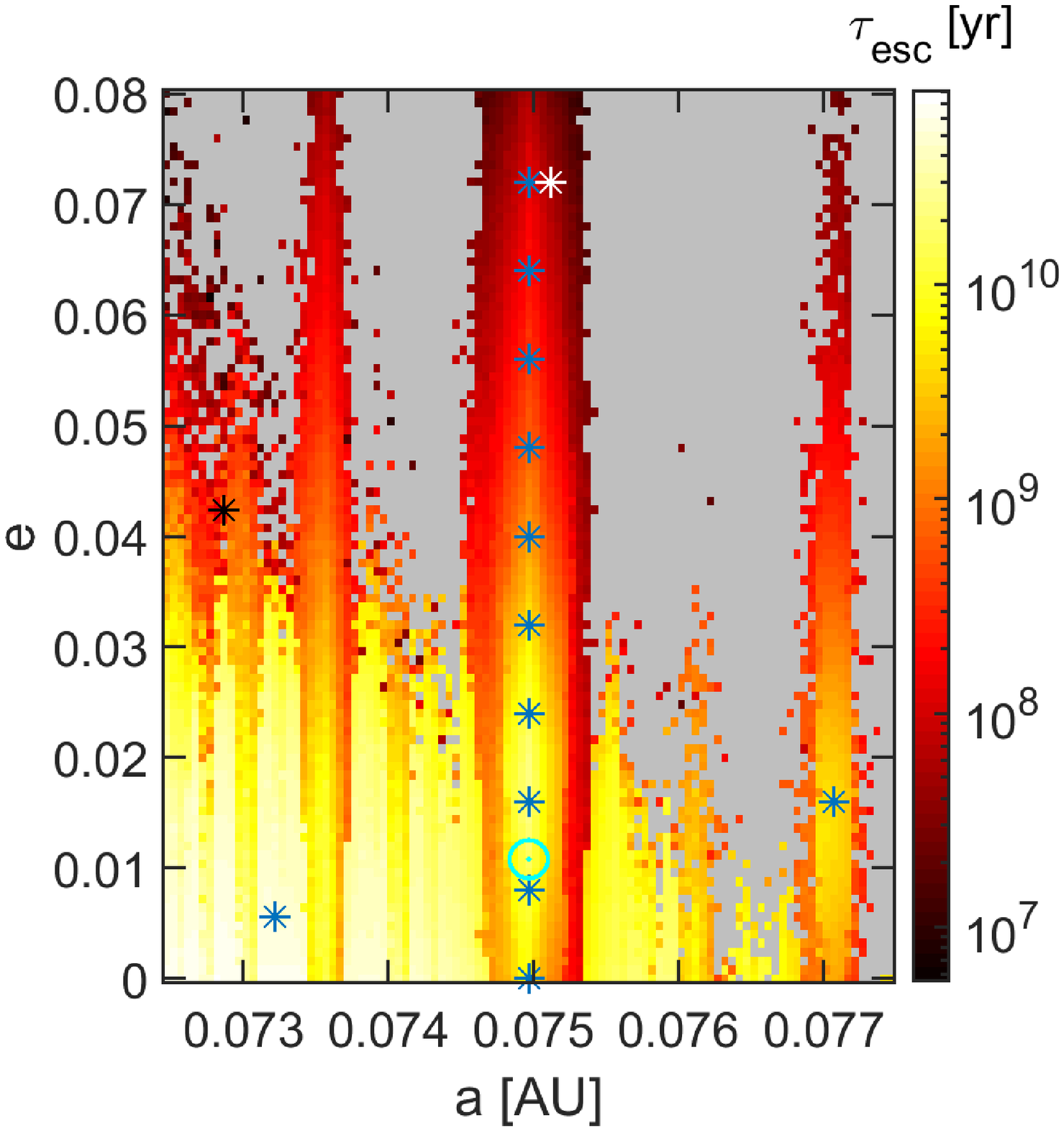}
	\end{subfigure}
	\caption{Dynamical maps for Kepler-60$b$, assuming a chain of $ 2 $-body resonances. The nominal position of planet~$ b $ is denoted with cyan-coloured circles at $ a_{b2} = 0.07497 $~AU, $ e_{b2} = 0.0108 $. The grey-coloured points did not reach the end of integration, i.e., with our previous denomination, they are not 'regular'. \textbf{Left panel:} eccentricity variations throughout the integration. \textbf{Middle panel:} normalized Shannon entropy at the end of integration. \textbf{Right panel:} stability times derived from the diffusion coefficients. (The asterisk symbols mark the initial conditions of long-term, direct integrations (see Section~\ref{subsec:comparison_with_direct_long_term_integrations}).)}
    \label{fig:K60B2}
\end{figure*}

Figure~\ref{fig:K60B2} shows the dynamical maps for planet~$ b $, assuming that the criteria for the libration of the critical angles of the 5:4 and 4:3 mean-motion commensurabilites are both satisfied. The first observation in the figure - in comparison with Figure~\ref{fig:K60B1} - is that the darker-coloured regions (of the first two panels) that are associated with the most regular orbits are now shifted to lower eccentricities, along with the cyan-coloured circle denoting the nominal position of planet~$ b $ at $ a_{b2} = 0.07497 $~AU, $ e_{b2} = 0.0108 $. Also, the upper halves of the maps became predominantly vacant and contain only the grey-coloured points that are not considered 'regular' in our notion. However, beside the primary island of the 5:4 MMR at the centre of the three panels, one discovers two additional, strong stripes near $ a \sim 0.0735 $ and $ \sim 0.077 $~AU. These bands, as compared to the case of the pure Laplace resonance, became much more extended in the vertical direction. In contrast, those being more evolved in the Laplacian case show now a shrinkage in length. As for the values of the eccentricity variations, the normalized Shannon entropy, and the stability times in the proximity of the primary 5:4 MMR, considerably lower $ \Delta e \lesssim 0.02 - 0.03 $, $ S/\ln(r) \sim 0.45 $, and larger $ \tau_\mathrm{esc} \sim 10^9-10^{10} $~yrs are seen this time. It implies that if the resonant criteria are fulfilled in between the adjacent planets $ b $, $ c $ and $ c $, $ d $ (and not only for the three planets altogether), then the system can remain stable one order of magnitude longer.

\subsubsection{The phase space of planet~$ c $}
\label{subsubsec:the_phase_space_of_planet_c2}

\begin{figure*}
    \begin{subfigure}{0.33\textwidth}
		\includegraphics[width = \textwidth]{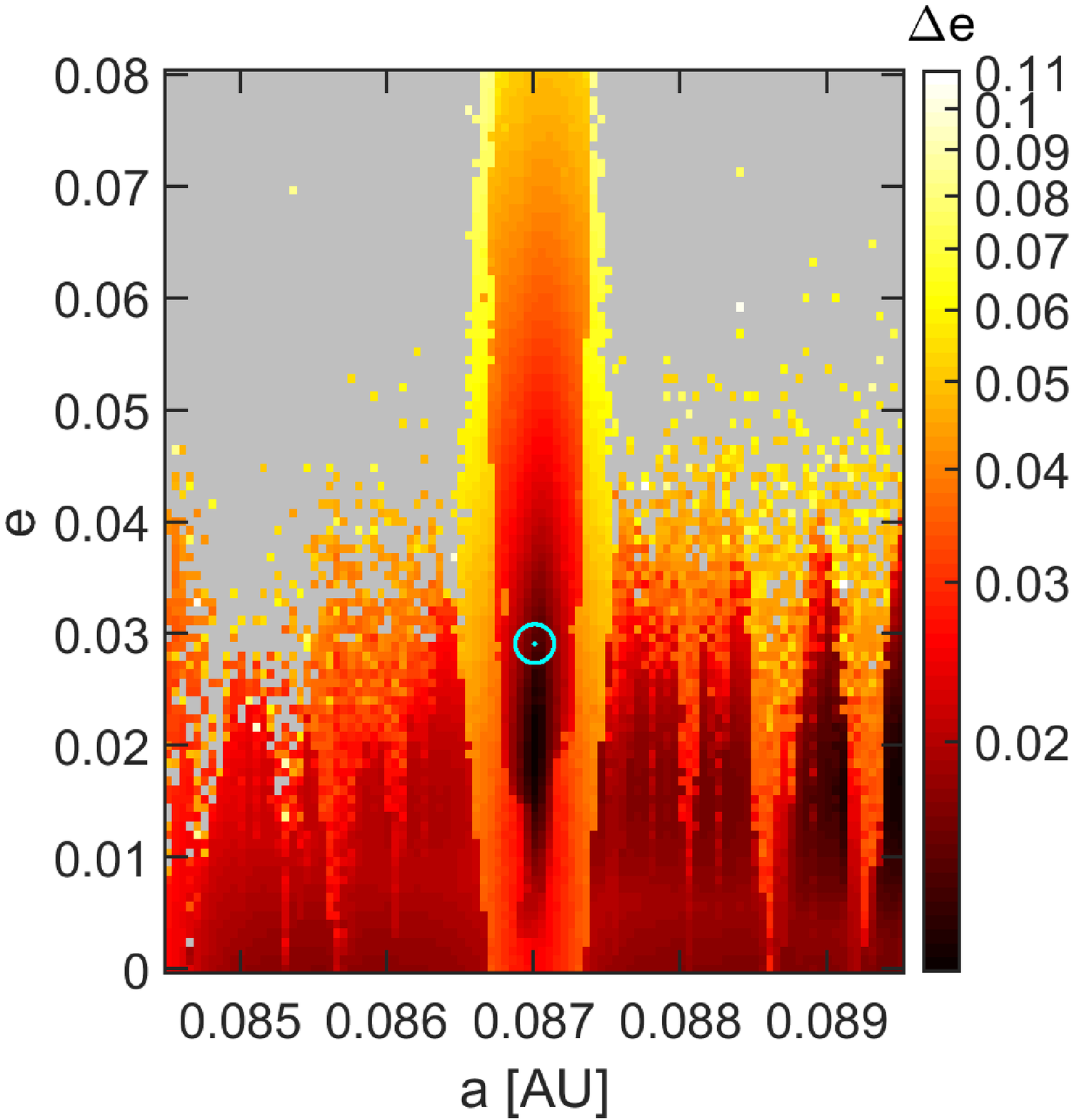}
	\end{subfigure}
	\begin{subfigure}{0.33\textwidth}
		\includegraphics[width = \textwidth]{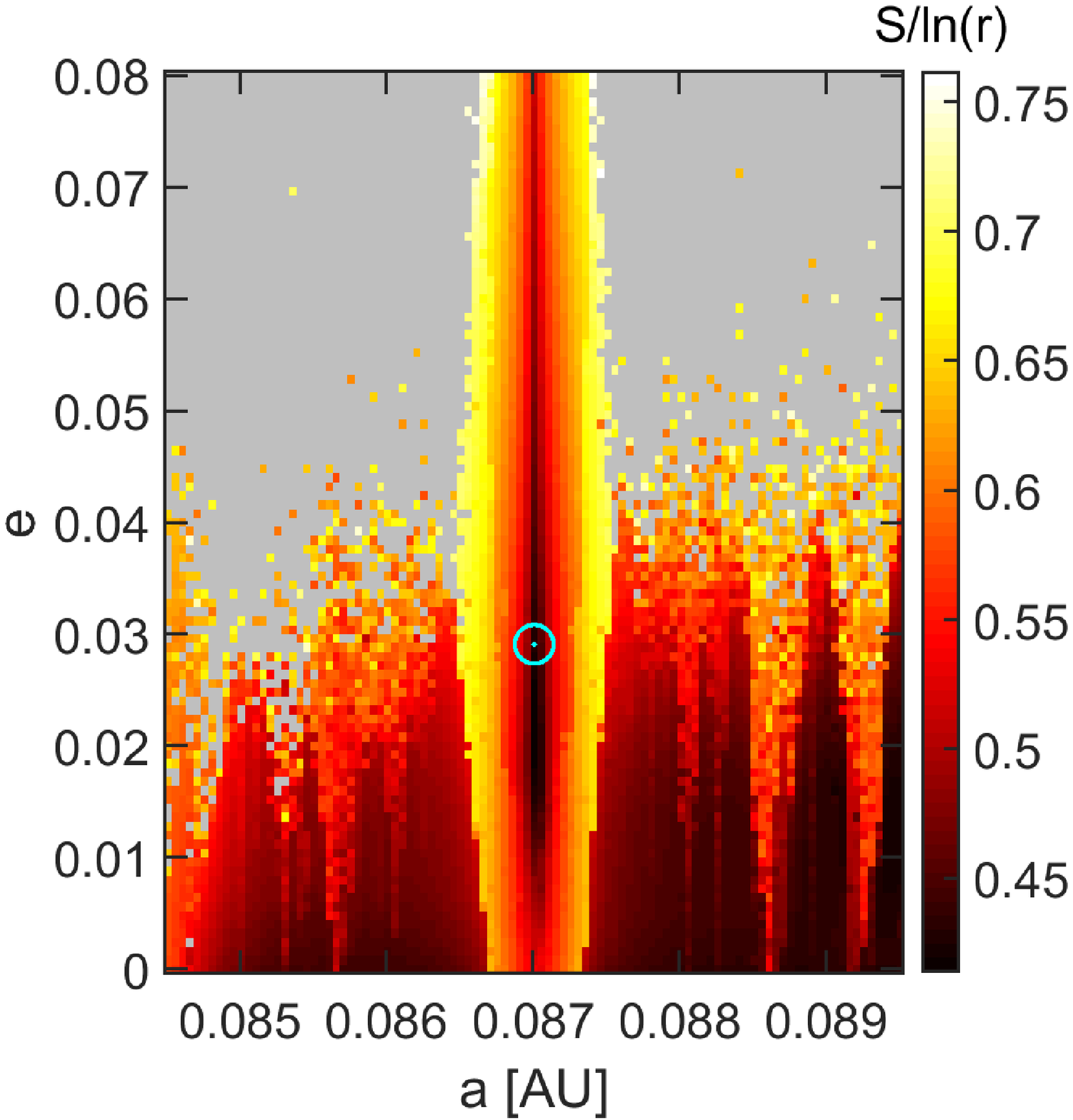}
	\end{subfigure}
	\begin{subfigure}{0.33\textwidth}
		\includegraphics[width = \textwidth]{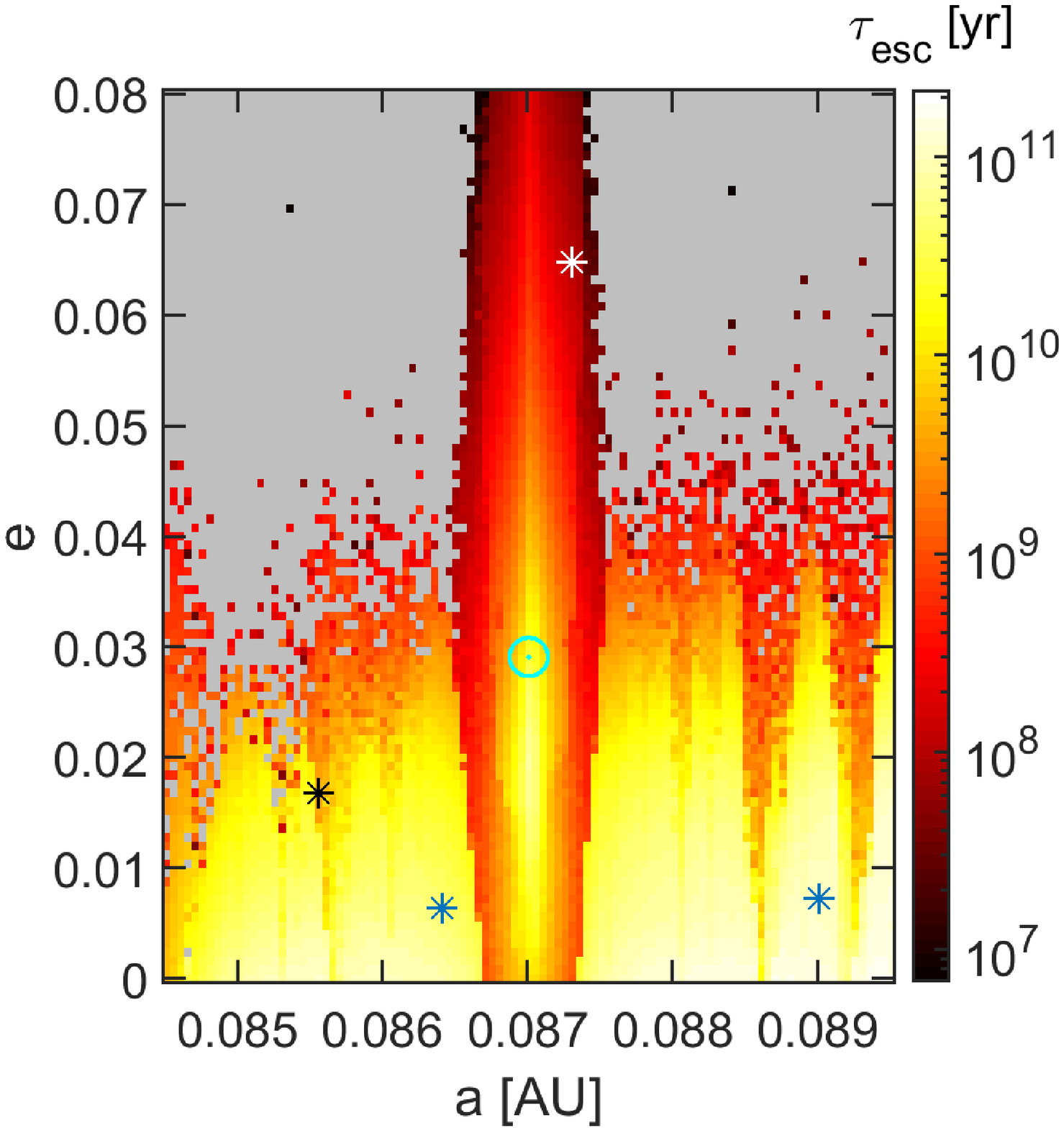}
	\end{subfigure}
	\caption{Dynamical maps for Kepler-60$c$, assuming a chain of $ 2 $-body resonances. The nominal position of planet~$ c $ is denoted with cyan-coloured circles at $ a_{c2} = 0.08701 $~AU, $ e_{c2} = 0.0291 $. The grey-coloured points did not reach the end of integration, i.e., with our previous denomination, they are not 'regular'. \textbf{Left panel:} eccentricity variations throughout the integration. \textbf{Middle panel:} normalized Shannon entropy at the end of integration. \textbf{Right panel:} stability times derived from the diffusion coefficients. (The asterisk symbols mark the initial conditions of long-term, direct integrations (see Section~\ref{subsec:comparison_with_direct_long_term_integrations}).)}
    \label{fig:K60C2}
\end{figure*}

According to Figure~\ref{fig:K60C2}, the same conclusions can be drawn in the case of the middle planet as in that of planet~$ b $: the most stable zones of the pure Laplacian case (see Figure~\ref{fig:K60C1}) are drifted near the bottom of the panels, while leaving the upper parts unoccupied by 'regular' orbits. The only exception is the primary island of the 2-body resonance at the centre. The neighbourhood of planet~$ c $ at the position $ a_{c2} = 0.08701 $~AU, $ e_{c2} = 0.0291 $, is characterized by $ \Delta e \lesssim 0.02 $, $ S/\ln(r) \sim 0.45 $, $ \tau_\mathrm{esc} \sim 10^{10} $~yrs.

\subsubsection{The phase space of planet~$ d $}
\label{subsubsec:the_phase_space_of_planet_d2}

\begin{figure*}
    \begin{subfigure}{0.33\textwidth}
		\includegraphics[width = \textwidth]{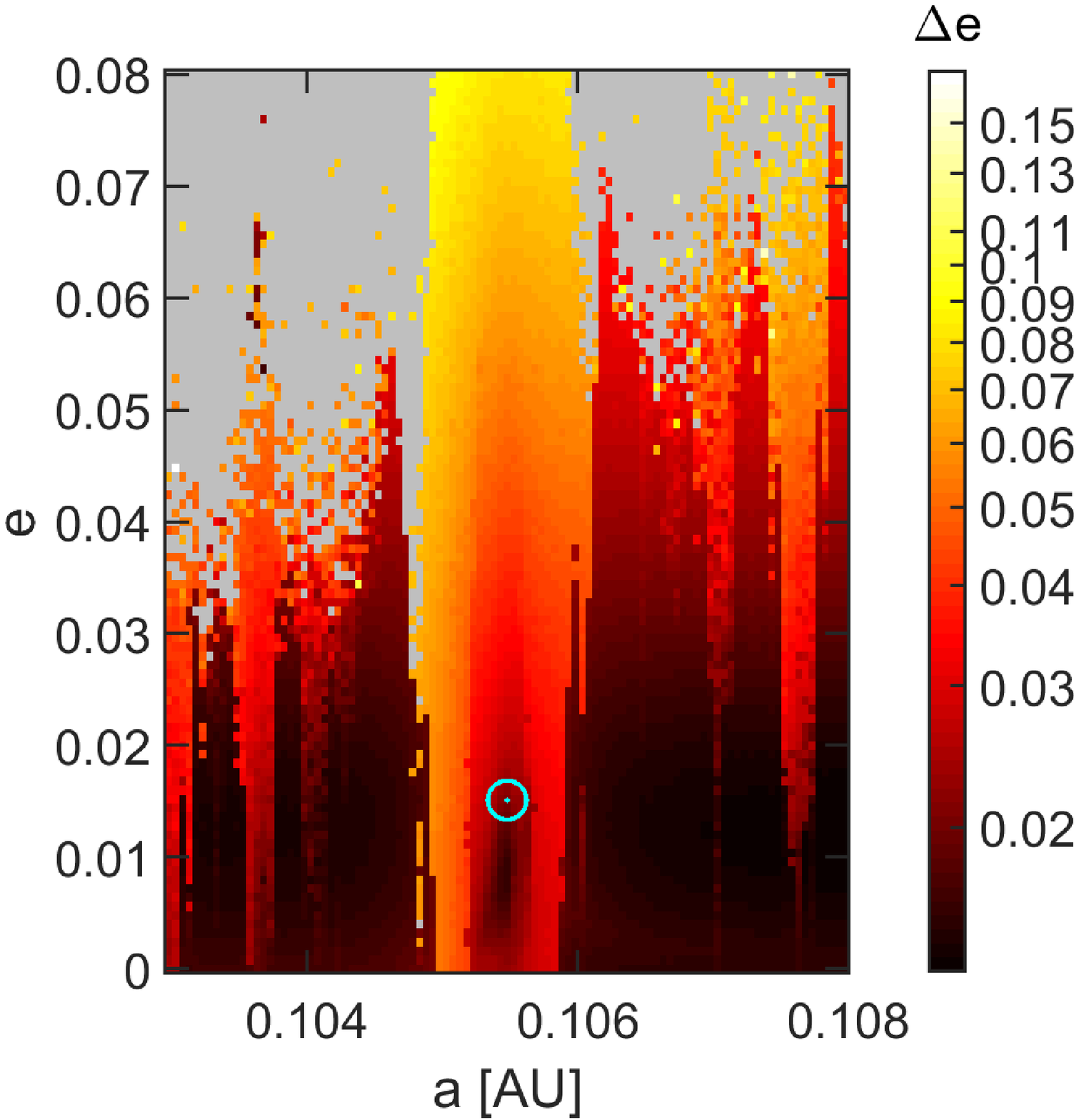}
	\end{subfigure}
	\begin{subfigure}{0.33\textwidth}
		\includegraphics[width = \textwidth]{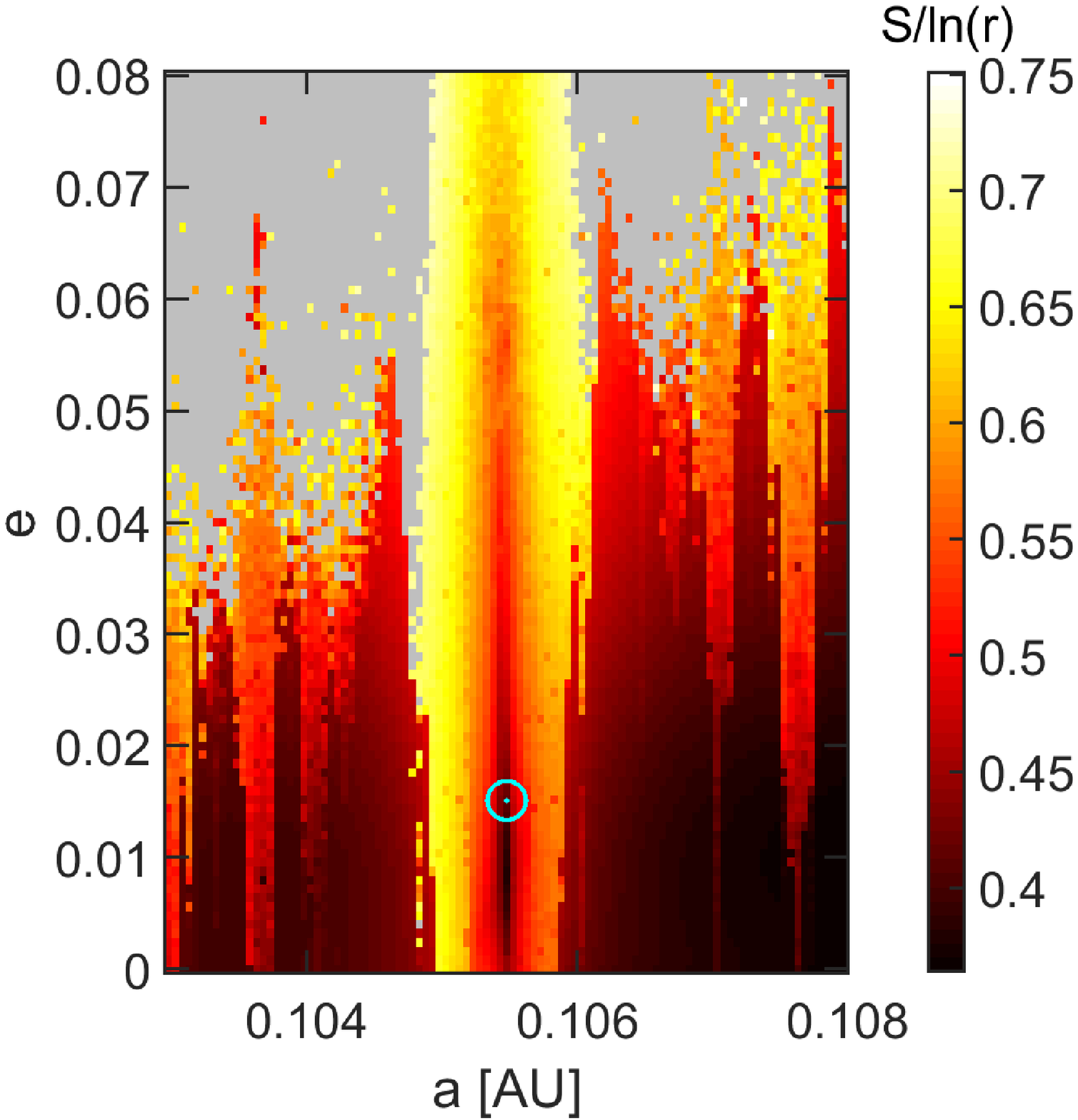}
	\end{subfigure}
	\begin{subfigure}{0.33\textwidth}
		\includegraphics[width = \textwidth]{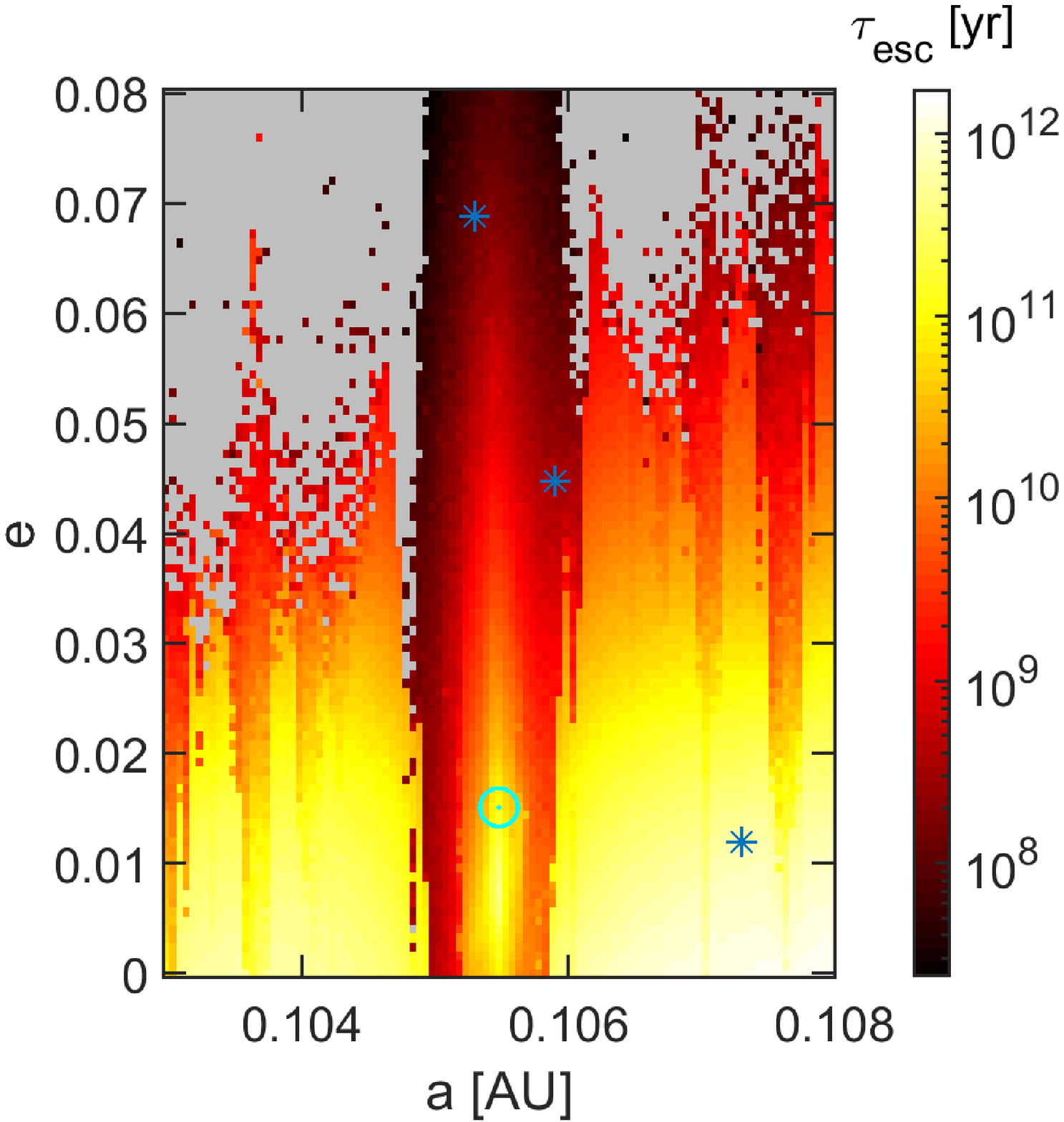}
	\end{subfigure}
	\caption{Dynamical maps for Kepler-60$d$, assuming a chain of $ 2 $-body resonances. The nominal position of planet~$ d $ is denoted with cyan-coloured circles at $ a_{d2} = 0.10548 $~AU, $ e_{d2} = 0.0151 $. The grey-coloured points did not reach the end of integration, i.e., with our previous denomination, they are not 'regular'. \textbf{Left panel:} eccentricity variations throughout the integration. \textbf{Middle panel:} normalized Shannon entropy at the end of integration. \textbf{Right panel:} stability times derived from the diffusion coefficients. (The asterisk symbols mark the initial conditions of long-term, direct integrations (see Section~\ref{subsec:comparison_with_direct_long_term_integrations}).)}
    \label{fig:K60D2}
\end{figure*}

The dynamical maps of planet~$ d $ in the case of the two $ 2 $-body resonances are presented in Figure~\ref{fig:K60D2}, but again, let us carry out our discussion by comparing the two cases of the resonance, i.e., see also Figure~\ref{fig:K60D1}. The overall relocation of the most stable realms in the present model of resonant configuration is apparent again, as is the filling of the gaps of the grey-coloured, non-'regular' points in between the resonant stripes of the Laplacian model. (The latter finding was even more accentuated in the case of the two inner planets, see the comparison of Figures~\ref{fig:K60B1} and \ref{fig:K60B2}, and Figures~\ref{fig:K60C1} and \ref{fig:K60C2}.) The characteristic values in the vicinity of planet~$ d $ (see the cyan circles at $ a_{d2} = 0.10548 $~AU, $ e_{d2} = 0.0151 $) are as follows: $ \Delta e \lesssim 0.02-0.03 $, $ S/\ln(r) \sim 0.45 $, $ \tau_\mathrm{esc} \gtrsim 10^{10} $~yrs.

\subsection{Comparison with direct, long-term integrations}
\label{subsec:comparison_with_direct_long_term_integrations}

\begin{figure*}
    \begin{subfigure}{0.33\textwidth}
		\includegraphics[width = \textwidth]{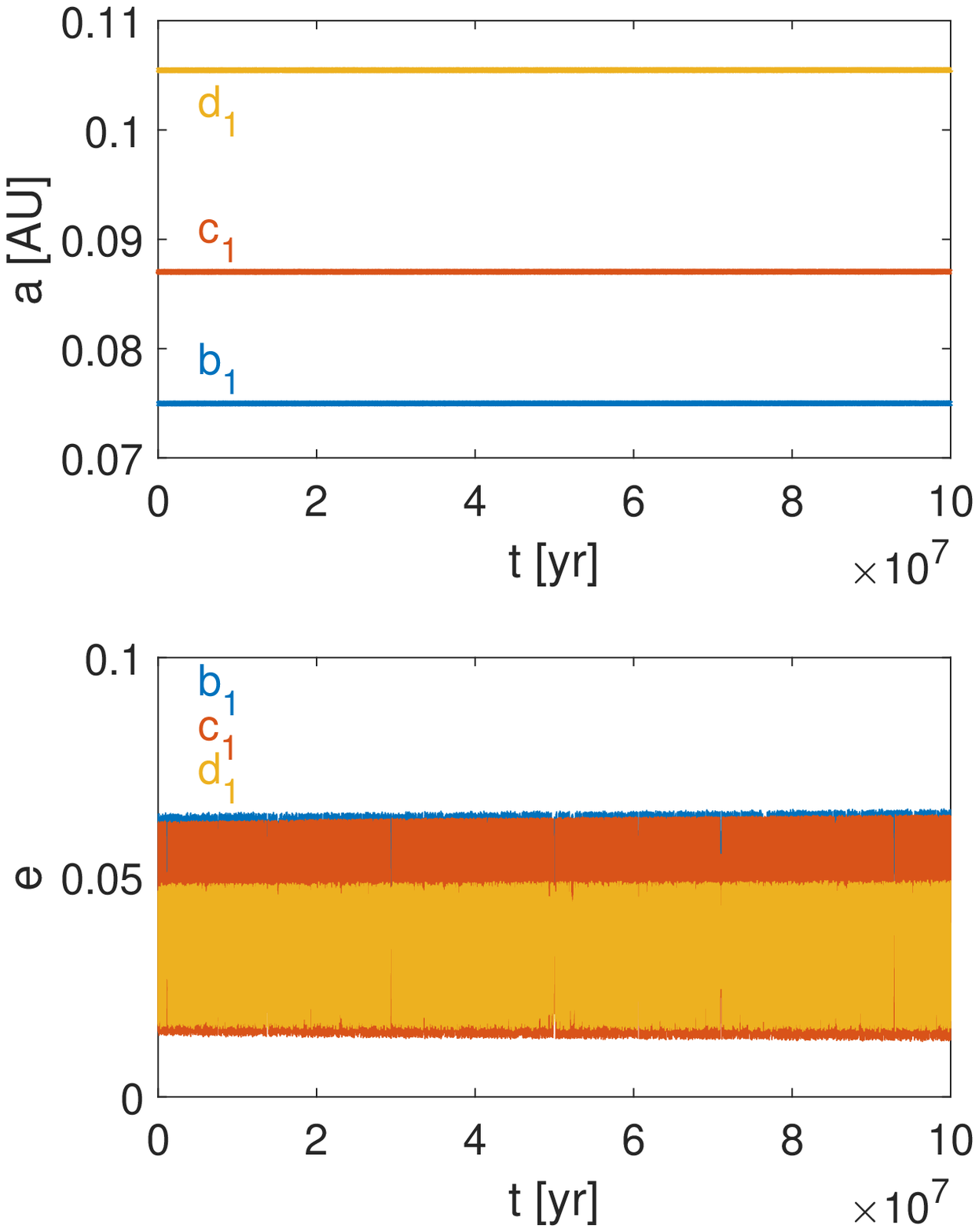}
	\end{subfigure}
	\qquad
	\begin{subfigure}{0.33\textwidth}
		\includegraphics[width = \textwidth]{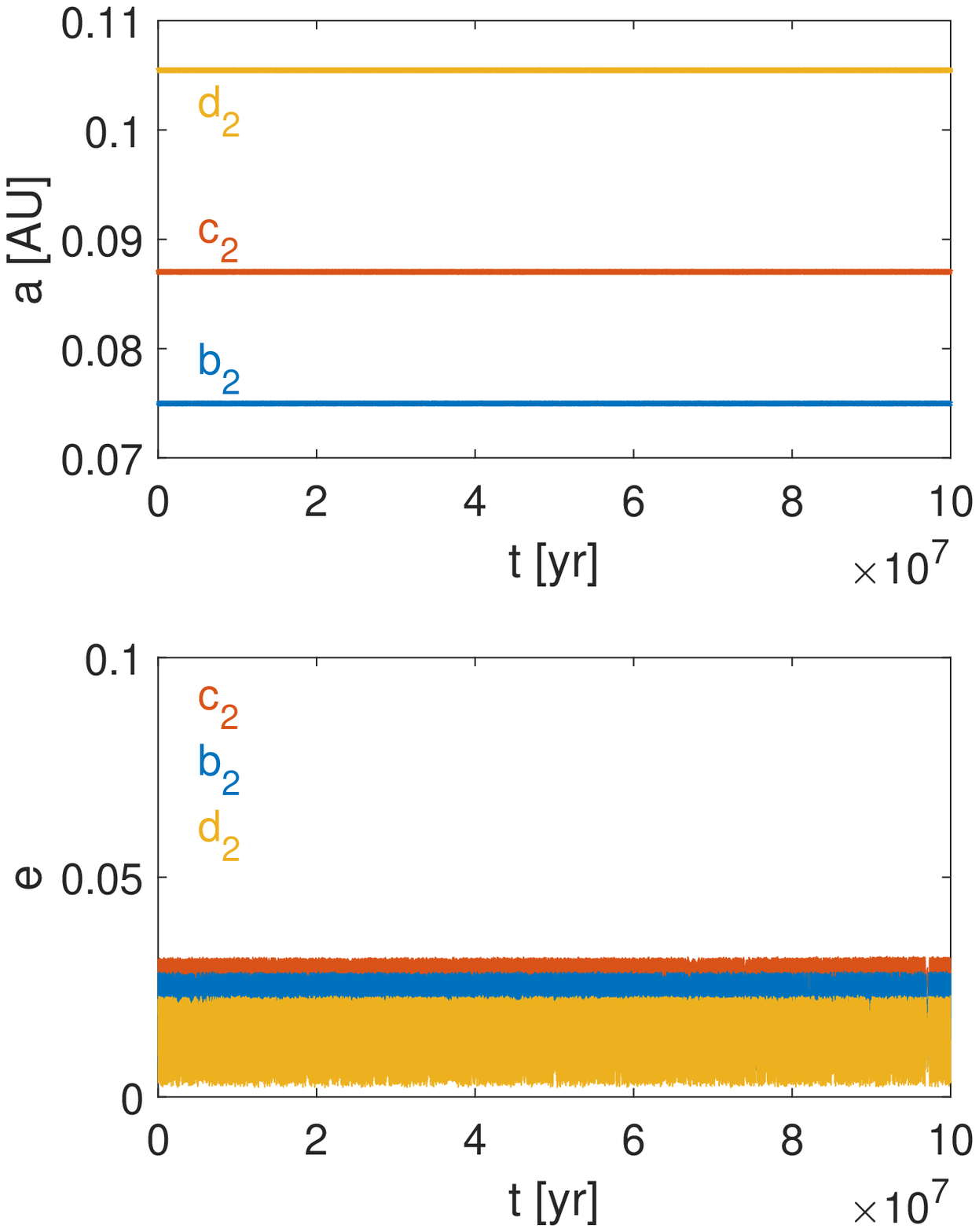}
	\end{subfigure}
	\caption{Time evolution of the semi-major axes and eccentricities of the nominal positions of the three planets (marked with the colours blue, russet, and ochre, respectively). Total integration time: $ 10^8 $~yrs, sampling timestep: 1000~yrs. \textbf{Left panels:} the case of the pure Laplace resonance. \textbf{Right panels:} the case of the chain of two 2-body resonances.}
    \label{fig:evol}
\end{figure*}

In the previous subsections, the stabilizing role of both the pure 3-body Laplace resonance and the chain of 2-body resonances became clear. We saw the safely wide primary islands of the above MMRs to surround the nominal positions of the planets, but also observed long-termly stable orbits in additional stripes of resonances of different types or of higher orders. The stability times deduced from the Shannon entropy yielded values as high as $ 10^{8}-10^{10} $~yrs.

In this subsection, we add the results of direct numerical integrations to our work, in order to have an independent confirmation of the long escape times.

Apart from the two nominal models, we ran 52 integrations altogether, placed at different initial locations in the $ (a, e) $ plane (see the asterisk symbols in the $ \tau_\mathrm{esc} $ maps of Figures~\ref{fig:K60B1}, \ref{fig:K60C1}--\ref{fig:K60D2}). With a Bulirsch--Stoer routine, we integrated the equations of motion for $ 10^8 $~yrs. The sampling timestep was chosen to be 1000~yrs.

The results for the nominal positions are seen in Figure~\ref{fig:evol} (the planets $ b $, $ c $, and $ d $ are marked with the colours blue, russet, and ochre, respectively). The time evolution of the semi-major axes as well as that of the eccentricities bear witness of great stability. The librational amplitudes of both $ a $ and $ e $ remain low during the whole time span. This fact foreshadows stability, in both cases of the resonance, to hold for 1 or 2 additional orders of time, in accordance with the escape times $ \tau_\mathrm{esc} $ obtained from the Shannon entropy method. What imply that the second model of the chain of two 2-body resonances is still favourable in terms of long-term stability are, on the one hand, its slightly smaller amplitudes in $ e $ and the more moderate mean values (also of $ e $) on the other.

The simulations of the further ICs show good agreement with the Shannon entropy results, too: the majority of these points (blue-coloured asterisks in Figures~\ref{fig:K60B1}, \ref{fig:K60C1}--\ref{fig:K60D2}) remained stable throughout the $ 10^8 $~yrs of integration or became unstable just prior to the end. We note, however, that among the $ 52 (+2) $ initial conditions integrated directly, we found 4 (see the black-coloured asterisks in Figures~\ref{fig:K60B1}, \ref{fig:K60C1}, \ref{fig:K60B2}, \ref{fig:K60C2}) where the corresponding system was disrupted after a few times $ 10^5 $~yrs. These occurrences of instability were not foretold by the entropy. Conversely, 5 of the ICs (white-coloured asterisks in Figures~\ref{fig:K60B1}, \ref{fig:K60C1}--\ref{fig:K60C2}) were chosen so that their escape times derived from the entropy were a few times $ 10^7 $~yrs ($ < 10^8 $~yrs). As for these points, the direct integrations revealed only an increment (or irregular variations) in the oscillation amplitudes of the actions $ a $ and $ e $, but the systems were not completely disrupted prior to $ 10^8 $~yrs. As regards the possible explanations of such irregularities within the 'regular', non-grey regions of the phase space, first, one observes that these specific, black- and white-coloured ICs are located close either to the boundaries of the 'regular' domains or to the separatrices of the MMRs. The Kepler-60 system is a very closely-packed one with its three massive and short-period planets, thus it becomes particularly sensitive to the conditions of computations\footnote{For the long-term integrations, instead of the MERCURY $ n $-body integrator, we used a code of our own.} near these borderlines. Furthermore, we did not use ensembles when calculating the Shannon entropy of single ICs, and as stated by \cite{cincotta2021febr} such extension of the calculations might increase the accuracy of the method. In our case, however, the latter upgrade would not have been feasible computationally, for the several cases considered and for the large number of ICs involved within each case to study not only the dynamics of the nominal states of the two models but the phase space around the planets, too. Yet the above findings concern only a limited fraction of the points, therefore we conclude that, statistically, the Shannon entropy approach is properly applicable in the four-body problem, and also, that for the two nominal positions the stability times deduced here are credible.

\section{Summary}
\label{sec:summary}

In this paper, we demonstrated how the Shannon entropy can be applied to investigate the dynamics of a resonant planetary system of four massive bodies (including the star, too).

The general structure of the phase space of a dynamical system is usually studied by using the classical chaos indicators such as the Lyapunov Characteristic Number \citep[LCN;][]{benettin1980}, the Mean Exponential Growth factor of Nearby Orbits \citep[MEGNO;][]{cincotta2000,cincotta2003}, the Fast Lyapunov Indicator \citep[FLI;][]{froeschle1997,guzzo2002}, the Relative Lyapunov Indicator \citep[RLI;][]{sandor2000,sandor2004}, and so on. These methods suffice for (quickly) detecting the most important resonances, chaotic regions, and islands of stability in the phase space; however, the deeper characteristics of such phenomena remain unrevealed by them. For instance, they are unable to distinguish between the stable and unstable chaos. The rate of the chaotic diffusion in the regions of irregular motion can not be directly derived by them either. Yet the latter quantity, for example, is of fundamental importance in understanding the overall and long-term dynamics of a given celestial system.

This is why the recent applications of the Shannon entropy bear great significance. On the one hand, the entropy serves as a reliable chaos indicator, by measuring the volume of the phase space that a trajectory of a single initial condition occupies during its time evolution. Moreover, in the case of (nearly) normal diffusion, the diffusion rate can also be determined in rather short integration times. The inverse of the diffusion coefficient then approximates the characteristic time of stability, thus one of the most fundamental questions related to a dynamical system can be answered quantitatively: how long the system will last.

In recent years, several applications of the Shannon entropy were introduced; however, we claim that the present paper is the first to test the entropy technique in the case of a four-body planetary system.

The Kepler-60 extrasolar system is a particularly interesting one, with its three super-Earth-sized planets engaged in a chain of mean-motion commensurabilities. Previous studies of the system \citep{papaloizou2015,gozdziewski2016} already drew attention to the ambiguous nature of the resonant dynamics. While mean-motion resonances assuredly play an important role in shaping the dynamics of the planets, the exact type of the resonant configuration was yet unclear. \cite{gozdziewski2016} proposed two possible scenarios to describe the resonant structure of the system. In their first suggestion the planets are involved in a pure Laplace resonance of the ratios 5:4:3 where the critical angles of the $ 2 $-body resonances 5:4 and 4:3 circulate and only that of the $ 3 $-body resonance librates. The second proposition is that all the critical angles librate thus both the $ 2 $-body resonances and the $ 3 $-body resonance are present.

Our aim was to map the phase space in the proximity of the three planets and inspect whether - by means of long-term stability - one solution is favoured over the other.

We constituted dynamical maps based on the Shannon entropy and its time derivative and also that of the eccentricity variations for a cross-check. Our results indicate that although extended regions of chaotic motion appear in some parts of the phase space, the resonances stabilize the configuration to a large extent and the planets are found in safely wide stable zones in the case of the pure Laplace resonance and in the case of the chain of $ 2 $-body resonances likewise. The stability times of the planets do show some differences in the two cases, however. For the pure Laplace resonance, the longest stability times, reached in the centre of the resonance, were $ \sim 10^9 $~yrs, whereas for the chain of the $ 2 $-body resonances we obtained stability times one magnitude longer: $ \sim 10^{10} $~yrs.

We also performed direct, long-term numerical integrations in the case of 54 initial conditions in order to have an independent verification of the indirectly derived escape times. The results of these simulations are in agreement with the long stability times deduced by means of the Shannon entropy approach.

The above findings suggest that the preferred configuration of the planets is the one in which the resonant criteria are fulfilled in between all adjacent bodies as well as for all three of the planets (i.e. the chain of two $ 2 $-body resonances). Our conclusions are in accordance with the propositions of \cite{gozdziewski2016} who stated that considering also the past evolution of the system, the chain of $ 2 $-body commensurabilities is the more probable outcome of a presumable convergent migration.

\section*{Acknowledgements}
This work was partly supported by the ÚNKP-20-3 and ÚNKP-21-3 New National Excellence Programs of the Ministry for Innovation and Technology from the source of the National Research, Development, and Innovation Fund.

EK and ZsS also acknowledge the support of the bilateral German--Hungarian Project CSITI (grant No.: 308019) financed by the DAAD and by the Tempus Public Foundation. ZsS thanks the support of the Hungarian National Research, Development, and Innovation Office (NKFIH) under the grant K-119993.

The authors thank, furthermore, Pablo Cincotta for his professional advice; Róbert Teravágimov for his valuable help in code designing; and the reviewer for the supportive comments and suggestions that helped us improve the manuscript.

\section*{Data Availability}

The data underlying this article will be shared on reasonable request to the corresponding author.



\bibliographystyle{mnras}
\bibliography{ref}

\bsp	
\label{lastpage}
\end{document}